\documentclass[12pt,eadjoint tfnpsf]{article}
\usepackage{footnote}

\catcode`\@=11
\@addtoreset{equation}{section}

\global\arraycolsep=1pt

\usepackage{amsbsy,amssymb,latexsym,amsfonts, amsmath}
\usepackage{mathrsfs}
\usepackage{graphicx}
\usepackage{youngtab}
\usepackage{marvosym}

\newcommand{\beq}{\begin{equation}}
\newcommand{\eeq}{\end{equation}}
\newcommand{\bea}{\begin{eqnarray}}
\newcommand{\eea}{\end{eqnarray}}

\def\E{{\epsilon_{1}}}
\def\EE{{\epsilon_{2}}}
\def\EEE{{\epsilon_3}}
\newcommand\tr{\mathrm{tr}}

\renewcommand{\thefootnote}{\fnsymbol{footnote}}

\def\XXint#1#2#3{{\setbox0=\hbox{$#1{#2#3}{\int}$}
     \vcenter{\hbox{$#2#3$}}\kern-.5\wd0}}

\begin{document}
%
%
\begin{titlepage}

\begin{flushright}
\normalsize
~~~~
SISSA  18/2014/FISI-MATE
\end{flushright}

\vspace{80pt}

\begin{center}
{\LARGE Six-dimensional supersymmetric gauge theories, \\ quantum cohomology of instanton moduli spaces \\ and \\ \vspace{0.5 cm}} {\LARGE gl(N) Quantum Intermediate Long Wave Hydrodynamics}
\end{center}

\vspace{25pt}

\begin{center}
{
Giulio Bonelli$^{\heartsuit\spadesuit}$, Antonio Sciarappa$^{\heartsuit}$, Alessandro Tanzini$^{\heartsuit}$ and Petr Vasko$^{\heartsuit}$
}\\
%
\vspace{15pt}
%
$^{\heartsuit}$
International School of Advanced Studies (SISSA) \\via Bonomea 265, 34136 Trieste, Italy 
and INFN, Sezione di Trieste \footnote{email: bonelli,asciara,tanzini,vaskop@sissa.it}\\
\vspace{15pt}
$^{\spadesuit}$
I.C.T.P.\\ Strada Costiera 11, 34014 Trieste, Italy
\end{center}
%
\vspace{20pt}
%

We show that the exact partition function of $U(N)$ six-dimensional gauge
theory with eight supercharges on $\mathbb{C}^2\times S^2$ provides the quantization
of the integrable system of hydrodynamic type known as $gl(N)$ periodic Intermediate Long Wave (ILW).
We characterize this system as the hydrodynamic limit of elliptic
Calogero-Moser integrable system. We compute the Bethe equations from the effective gauged linear sigma model
on $S^2$ with target space the ADHM instanton moduli space, whose mirror computes the
Yang-Yang function of $gl(N)$ ILW.  
The quantum Hamiltonians are given by the local chiral ring observables of the six-dimensional
gauge theory. As particular cases, these provide the $gl(N)$ Benjamin-Ono and Korteweg-de Vries quantum
Hamiltonians. In the four dimensional limit, we identify the local chiral ring observables 
with
the conserved charges of Heisenberg plus $W_N$ algebrae, thus providing a gauge theoretical proof
of AGT correspondence.  
   


\vfill

\setcounter{footnote}{0}
\renewcommand{\thefootnote}{\arabic{footnote}}

\end{titlepage}

\tableofcontents

\section{Introduction}
\label{sec:intro}

The intimate relation among BPS correlators of $\mathcal{N}=2$ supersymmetric gauge theories in four
dimensions, two-dimensional  
conformal field theories and 
integrable systems revealed 
a world of surprises 
behind the Seiberg-Witten solution 
of ${\cal N}=2$ supersymmetric 
$d=4$ gauge theories in the Coulomb branch \cite{Seiberg:1994rs}.
Its microscopic derivation via equivariant instanton counting \cite{Nekrasov:2002qd} 
pointed to a connection with free two-dimensional conformal field theories
\cite{Losev:2003py,Nekrasov:2003rj,2008arXiv0801.2565C}.
More recently the embedding in M-theory \cite{Witten:1997sc,Gaiotto:2009we}
paved the way to the realization of the AGT correspondence with Liouville and Toda theory
\cite{Alday:2009aq,Wyllard:2009hg}.

The relation between $d=4$ ${\cal N}=2$ gauge theories and integrable systems was understood to underlie
the ability of solving the effective theory in the IR already in the nineties in the context of the Seiberg-Witten theory
\cite{Gorsky:1995zq,Donagi:1995cf,Martinec:1995qn,D'Hoker:1997ha}, 
the prototypical example being the Toda lattice. 

In this paper we discuss a new connection between supersymmetric theories with eight supercharges and quantum integrable systems of {\it hydrodynamical} type.
These naturally arise in the context of AGT correspondence.
Indeed integrable systems and conformal field theories in two dimensions are intimately connected from several 
points of view.
The link between conformal field theory and quantum KdV was noticed in 
\cite{Sasaki:1987mm,Eguchi:1989hs,Kupershmidt:1989bf,Bazhanov:1994ft}.
In \cite{Bazhanov:1994ft}
the infinite conserved currents in involutions of the Virasoro algebra $Vir$ 
have been shown to realize the quantization of the KdV system
and the quantum monodromy ``T-operators'' are shown to act on 
highest weight Virasoro modules.

More recently an analogous connection between the spectrum of a CFT based on the
Heisenberg plus Virasoro algebra $H\oplus Vir$ and the bidirectional Benjamin-Ono (BO${}_2$)
system has been shown in the context of a combinatorial proof of AGT correspondence
\cite{Alba:2010qc}, providing a first example of the phenomenon we alluded to before.

In \cite{Bonelli:2013rja} the exact partition function of the six dimensional $U(N)$ 
supersymmetric gauge theory on $S^2\times\mathbb{C}^2$ was computed and in particular shown to 
account for the $S^2$-finite size corrections to the Nekrasov partition function.
From a mathematical viewpoint these corrections compute the quantum cohomology of the ADHM moduli space of 
instantons in terms of quasi-map ${\cal I}$ and ${\cal J}$ functions,
the complexified K\"ahler parameter being identified with the Fayet-Iliopoulos (FI) parameter
of the effective Gauged Linear Sigma Model (GLSM) on $S^2$.

In this paper we study the link between the six dimensional $U(N)$ exact partition function and 
quantum integrable systems finding that the supersymmetric gauge theory provides the quantization 
of the $gl(N)$ Intermediate Long Wave system (ILW${}_N$).
This is a well known one parameter deformation of the BO system. Remarkably, it interpolates 
between BO and KdV.
We identify the deformation parameter with the FI of the $S^2$ GLSM, by matching
the twisted superpotential of the GLSM
with 
the Yang-Yang function of quantum ILW${}_N$ as proposed in \cite{Litvinov:2013zda}.
Our result shows that the quantum cohomology of the ADHM instanton moduli space is 
computed by the quantum ILW${}_N$ system.
In the abelian case $N=1$, when the ADHM moduli space reduces to the Hilbert scheme of points on 
$\mathbb{C}^2$, this correspondence is discussed in \cite{Otalk,Ntalk,NOinp}.

On top of this we show that the chiral ring observables of the six dimensional gauge theory
are related to the commuting quantum Hamiltonians of ILW${}_N$.

Let us remark that in the four dimensional limit our results imply that the gauge theory chiral ring 
provides a basis for the BO${}_N$ quantum Hamiltonians. 
This shows the appearance of the $H\oplus { W}_N$
algebra in the characterization of the BPS sector of the four dimensional gauge theory as
proposed in \cite{2010PhLB..691..111B} and is a strong purely gauge theoretic argument in 
favour of the AGT correspondence.

We also show that classical ILW hydrodynamic equations arise as a collective description
of elliptic Calogero-Moser integrable system.
Let us notice that the quantum integrability of the BO${}_N$ system can be shown by constructing
its quantum Hamiltonians in terms of $N$ copies of trigonometric Calogero-Sutherland Hamiltonians
with tridiagonal coupling: a general proof in the context of equivariant quantum cohomology of 
Nakajima quiver varieties can be found in \cite{Maulik:2012wi}. The relevance of this construction in
the study of conformal blocks of W-algebra is discussed in \cite{Estienne:2011qk}. Our result hints to an analogous r\^ole
of elliptic Calogero system in the problem of the quantization of ILW${}_N$.
  
It is worth to remark at this point that these quantum systems play a relevant r\^ole in the description
of Fractional Quantum Hall liquids. In particular our results suggest the quantum ILW system to be useful 
in the theoretical investigation of FQH states on the torus, which are also more amenable to numerical
simulations due to the periodic boundary conditions. For a discussion on quiver gauge theories and
FQHE in the context of AGT correspondence see \cite{Santachiara:2010bt,Estienne:2011qk}.

This paper is organized as follows. In Section 2 we first recall some basic notions on the relevant integrable systems
and then discuss the ILW equations as hydrodynamical limit of elliptic Calogero-Moser.
In Section 3 we recall the results of \cite{Bonelli:2013rja} on the exact partition function of the six dimensional
$U(N)$ gauge theory on $\mathbb{C}^2\times S^2$ and their relation with the equivariant cohomology of the ADHM instanton
moduli space. In Section 4 we discuss the Landau-Ginzburg mirror and its relation with quantum ILW system and its KdV limit.
Section 5 is left for concluding remarks and discussions on open problems.
 
\section{Intermediate Long Wave system}
\label{ILW}

In subsection \ref{P} we recall some basic facts about $gl(N)$ ILW integrable hydrodynamics
which are relevant for the comparison with the six dimensional $U(N)$ gauge theory.
In the subsequent subsection \ref{eCS} we show that ILW system can be obtained as hydrodynamical limit of
elliptic Calogero-Moser system. 

\subsection{The prequel}
\label{P}

One of the most popular integrable systems is the KdV equation
\beq
u_t=2u u_x+\frac{\delta}{3} u_{xxx}
\label{kdv}\eeq
where $u=u(x,t)$ is a real function of two variables.
It describes the surface dynamics of shallow water in a channel,
$\delta$ being the dispersion parameter.

The KdV equation is a particular case of the ILW equation
\beq
u_t=2 u u_x+\frac{1}{\delta}u_x+{\cal T}[ u_{xx}]
\label{ilw}\eeq
where
${\cal T}$ is the integral operator
\beq
{\cal T}[f](x)=P.V.\int{\rm coth}\left(\frac{\pi(x-y)}{2\delta}\right)f(y) \frac{dy}{2\delta}
\label{intop}\eeq
and $P.V.\int$ is the principal value integral

Equation (\ref{ilw}) describes the surface dynamics of water in a channel of finite depth. 
It reduces
to (\ref{kdv}) in the limit
of small $\delta$.
The opposite limit, that is the infinitely deep channel at $\delta\to\infty$, is called the Benjamin-Ono equation.
It reads
\beq
u_t=2 u u_x+ H[u_{xx}]
\label{bo1}\eeq
where $H$ is the integral operator implementing the Hilbert transform on the real line
\beq
H[f](x)=P.V.\int\frac{1}{x-y}f(y) \frac{dy}{\pi}
\label{hilbop}\eeq

The equation (\ref{ilw}) is an integrable deformation of KdV. It has been proved in 
\cite{0305-4470-37-32-L02}
that the form of the integral kernel in \eqref{intop} is fixed by the requirement of integrability.

The version of the ILW system which we will show to be relevant to our case is the periodic one.
This is obtained by replacing (\ref{intop}) with
\beq
{\cal T}[f](x)=\frac{1}{2\pi} P.V.\int_0^{2\pi}
\frac{\theta_1'}{\theta_1}\left(\frac{y-x}{2},q\right)f(y) dy
\label{perintop}\eeq
where $q=e^{-\delta}$.

Equation (\ref{ilw}) is Hamiltonian with respect to the Poisson bracket
\beq
\{u(x),u(y)\}=\delta'(x-y)
\label{1poisson}
\eeq
and reads
\beq
u_t(x)=\{I_3,u(x)\}
\label{hamilw}\eeq
where $I_3=\int \frac{1}{3} u^3 +\frac{1}{2} u {\cal T}[u_{x}]$ is the corresponding Hamiltonian.
The other flows are generated by $I_2=\int\frac{1}{2}u^2$ and the further Hamiltonians $I_n=\int \frac{1}{n}u^n+\ldots$, where $n>3$,
which are determined by the condition of being in involution $\{I_n,I_m\}=0$.
These have been computed explicitly in \cite{LR}. 
The $gl(N)$ ILW system is described in \cite{LR2}; more explicit formulae for the $gl(2)$ case can be found 
in Appendix A of \cite{Litvinov:2013zda}.

The periodic ILW system can be quantized by introducing creation/annihilation operators corresponding to the Fourier modes of the 
field $u$ and then by the explicit construction of the quantum analogue of the commuting Hamiltonians $I_n$ above.
Explicitly, one introduces the Fourier modes $\{\alpha_k\}_{k\in{\mathbb Z}}$ with commutation relations 
$$\left[\alpha_k,\alpha_l\right]=k\delta_{k+l}$$
and gets the first Hamiltonians as
$$I_2=2\sum_{k>0}\alpha_{-k}\alpha_k-\frac{1}{24},$$
\beq
I_3=-\sum_{k>0}k{\rm coth}(k\pi t)\alpha_{-k}\alpha_k+\frac{1}{3}\sum_{k+l+m=0}\alpha_k\alpha_l\alpha_m
\label{I3}\eeq
where we introduced a {\it complexified} ILW deformation parameter
$2\pi t= \delta - i \theta$. This arises naturally in comparing 
the Hamiltonian (\ref{I3}) with the deformation of the quantum trigonometric Calogero-Sutherland Hamiltonian
appearing in the study of the quantum cohomology of ${\rm Hilb}^n\left({\mathbb C}^2\right)$ \cite{OP,OP1}, see Appendix \ref{NB} for 
details.
We are thus led to identify the creation and annihilation operators of the quantum periodic ILW system with
the Nakajima operators describing the equivariant cohomology of the instanton moduli space:
this is the reason why one has to consider {\it periodic} ILW to make a comparison with gauge theory results.
Moreover, from \eqref{I3} the complexified deformation parameter of the ILW system 
$2\pi t= \delta - i \theta$ gets identified with the K\"ahler parameter of the Hilbert scheme of points as $q=e^{-2\pi t}$.
In this way
the quantum ILW hamiltonian structure reveal to be related to abelian six dimensional gauge theories via BPS/CFT 
correspondence. In particular the BO limit $t\to \pm\infty$ corresponds to the classical equivariant cohomology of the instanton
moduli space described by the four dimensional limit of the abelian gauge theory. 

More general integrable systems of similar type arise by considering richer symmetry structures.
These are related to non-abelian gauge theories.
A notable example is that of ${ H}\oplus { Vir}$, where ${ H}$ is the Heisenberg algebra of a single 
chiral $U(1)$ current.
Its integrable quantization depends on a parameter which weights how to couple the generators of the two algebras 
in the conserved Hamiltonians. 
The
 construction of the corresponding quantum ILW system can be found in \cite{Litvinov:2013zda}.
This quantum integrable system, in the BO$_2$ limit, has been shown in \cite{Alba:2010qc} to govern the AGT realization of the $SU(2)$ ${\cal N}=2$
$D=4$ gauge theory with $N_f=4$. More precisely, the expansion of the conformal blocks proposed in \cite{Alday:2009aq}
can be proved to be the basis of descendants in CFT which diagonalizes the BO$_2$ Hamiltonians.

More in general one can consider the algebra ${ H}\oplus { W_{N}}$.
The main aim of this paper is to show that 
the partition function of the non-abelian six-dimensional gauge theory on $S^2\times {\mathbb C}^2$
naturally computes such a quantum generalization. Indeed, as it will be shown in Section 4,  
the Yang-Yang function of this system, as it is described in \cite{Litvinov:2013zda}, arises as the twisted superpotential of 
the effective LG model governing the finite volume effects of the two-sphere.
In particular, we propose that the Fourier modes of the $gl(N)$ periodic ILW system correspond to the 
Baranovsky operators acting on the equivariant cohomology of the ADHM instanton moduli space. Evidence for this proposal is given
in Section 4 and in the Appendix \ref{NB}.
Moreover in Section 4 we identify the deformation parameter $t$ in \eqref{I3} 
with the FI parameter of the gauged linear 
sigma model on the two sphere.

This generalizes the link between quantum deformed Calogero-Sutherland system and the abelian gauge theory to 
the $gl(N)$ ILW quantum integrable system and the non abelian gauge theory in six dimensions.

\subsection{ILW as hydrodynamical limit of elliptic Calogero-Moser}
\label{eCS}

 An important property of the non-periodic ILW system is that its rational solutions are determined by the 
trigonometric Calogero-Sutherland model (see \cite{opac-b1120330} for details).
In this subsection we show a similar result for periodic ILW, namely that the dynamics of the poles of
multisoliton solutions for this system is described by elliptic Calogero-Moser.   
Analogous results were obtained in \cite{Abanov,Hone}.
We proceed by generalizing the approach
of \cite{Abanov:2008ft} where this limit was discussed for trigonometric Calogero-Sutherland
versus the BO equation. The strategy is the following: one studies multi-soliton solutions to the ILW
system by giving a pole ansatz. The dynamics of the position of the poles turns out to be described
by an auxiliary system equivalent to the eCM equations of motion in Hamiltonian formalism.  

The Hamiltonian of eCM  system for $N$ particles is defined as
\begin{equation}
H_{eCM}=\frac{1}{2}\sum_{j=1}^N p_j^2+G^2\sum_{i<j}\wp(x_i-x_j;\omega_1,\omega_2),
\end{equation}
where $\wp$ is the elliptic Weierstrass $\wp$-function and the periods are chosen as $2\omega_1=L$ and $2\omega_2=i\delta$. 
In the previous Section 2.1 and in Sections 4,5 we set $L=2\pi$.
For notational simplicity, from now on we suppress the periods in all elliptic functions. 
The Hamilton equations read
\begin{align}
\label{eq:HEoM}
\dot{x}_j&=p_j \notag \\
\dot{p}_j&=-G^2\partial_j\sum_{k\neq j}\wp(x_j-x_k),
\end{align}
which can be recast as a second order equation of motion
\begin{equation}
\label{eq:LEoM}
\ddot{x}_j=-G^2\partial_j\sum_{k\neq j}\wp(x_j-x_k).
\end{equation}
It can be shown (see the Appendix \ref{eCM} for a detailed derivation) that equation \eqref{eq:LEoM} is equivalent to the following auxiliary system 
\footnote{Actually, the requirement that this system should reduce to \eqref{eq:LEoM} is not sufficient to 
fix the form of the functions appearing. As will be clear from the derivation below, we could as well substitute $\frac{\theta_1^{\prime}\left(\frac{\pi}{L}z\right)}{\theta_1\left(\frac{\pi}{L}z\right)}$ by $\zeta(z)$ and the correct equation of motion would still follow. 
However, we can fix this freedom by taking the trigonometric limit ($\delta\to\infty$) and requiring that 
this system reduces to the one in \cite{Abanov:2008ft}.}
\begin{align}
\label{eq:auxsystem}
\dot{x}_j&=iG\Bigg\{\sum_{k=1}^N\frac{\theta_1^{\prime}\left(\frac{\pi}{L}(x_j-y_k)\right)}{\theta_1\left(\frac{\pi}{L}(x_j-y_k)\right)}-\sum_{k\neq j}\frac{\theta_1^{\prime}\left(\frac{\pi}{L}(x_j-x_k)\right)}{\theta_1\left(\frac{\pi}{L}(x_j-x_k)\right)} \Bigg\} \notag \\ 
\dot{y}_j&=-iG\Bigg\{\sum_{k=1}^N\frac{\theta_1^{\prime}\left(\frac{\pi}{L}(y_j-x_k)\right)}{\theta_1\left(\frac{\pi}{L}(y_j-x_k)\right)}-\sum_{k\neq j}\frac{\theta_1^{\prime}\left(\frac{\pi}{L}(y_j-y_k)\right)}{\theta_1\left(\frac{\pi}{L}(y_j-y_k)\right)} \Bigg\}.
\end{align}
In the limit  $\delta\to \infty$ ($q\to 0$), 
the equation of motion \eqref{eq:LEoM} reduces to
\begin{equation}
\label{eq:EoMtrig}
\ddot{x}_j=-G^2\left(\frac{\pi}{L}\right)^2\partial_j\sum_{k\neq j}\cot^2\left(\frac{\pi}{L}(x_j-x_k)\right),
\end{equation}
while the auxiliary system goes to
\begin{alignat}{3}
\label{eq:auxsystemtrig}
\dot{x}_j&=iG\frac{\pi}{L}\Bigg\{\sum_{k=1}^N\cot\left(\frac{\pi}{L}(x_j-y_k)
\right)-\sum_{k\neq j}\cot\left(\frac{\pi}{L}(x_j-x_k)
\right)\Bigg\} \notag \\
\dot{y}_j&=-iG\frac{\pi}{L}\Bigg\{\sum_{k=1}^N\cot\left(\frac{\pi}{L}(y_j-x_k)
\right)-\sum_{k\neq j}\cot\left(\frac{\pi}{L}(y_j-y_k)
\right)\Bigg\} 
\end{alignat}
This is precisely the form obtained in \cite{Abanov:2008ft}.

In analogy with \cite{Abanov:2008ft} we can define a pair of functions which encode particle positions 
as simple poles
\begin{align}
u_1(z)&=-iG\sum_{j=1}^N\frac{\theta_1^{\prime}\left(\frac{\pi}{L}(z-x_j)\right)}{\theta_1\left(\frac{\pi}{L}(z-x_j)\right)} \notag \\
u_0(z)&=iG\sum_{j=1}^N\frac{\theta_1^{\prime}\left(\frac{\pi}{L}(z-y_j)\right)}{\theta_1\left(\frac{\pi}{L}(z-y_j)\right)}
\end{align}
and we also introduce their linear combinations 
\begin{equation}
u=u_0+u_1, \;\; \widetilde{u}=u_0-u_1.
\end{equation}
These satisfy the differential equation
\begin{equation}
\label{eq:diffeq}
u_t+uu_z+i\frac{G}{2}\widetilde{u}_{zz}=0,
\end{equation}
as long as $x_j$ and $y_j$ are governed by the dynamical equations \eqref{eq:auxsystem}.
The details of the derivation can be found in the Appendix \ref{eCM}.
Notice that, when the lattice of periodicity is rectangular, \eqref{eq:diffeq} is nothing but
ILW equation. Indeed, under the condition $x_i=\bar y_i$ one can show that $\tilde u = -i \mathcal{T}u$
\cite{LR}. To recover \eqref{ilw} one has to further rescale $u\to G u$ and $t\to -t/G$ and shift
$u\to u+1/2\delta$.
We observe that \eqref{eq:diffeq} does not explicitly depend on the number of particles $N$ and holds
also in the hydrodynamical limit $N,L\to\infty$, with $N/L$ fixed.

\section{Partition function of ${\cal N}=1$ Super Yang-Mills theory on $\mathbb{C}^2\times S^2$}
\label{1}

The partition function of ${\cal N}=1$ Super Yang-Mills theory on $\mathbb{C}^2\times S^2$ with $U(N)$ gauge group
in presence of $\Omega$-background
was computed in \cite{Bonelli:2013rja}. It is given by the product of a 1-loop term and a non perturbative contribution, namely
\beq
{\cal Z}={\cal Z}_{1-{\rm loop}}{\cal Z}_{{\rm np}}
\label{Z6}\eeq
where
\beq
{\cal Z}_{1-{\rm loop}}=\prod_{l\neq m}
\Gamma_2(a_{lm};\E,\EE)\frac{\Gamma_3\left(a_{lm};\E,\EE,\frac{1}{ir}\right)}
{\Gamma_3\left(a_{lm};\E,\EE,-\frac{1}{ir}\right)}
\label{Z61l}\eeq
\beq
{\cal Z}_{{\rm np}}=\sum_{k\geq 0}Q^k {\cal Z}_k\left(
\vec a,\E,\EE;q,\bar q,r
\right)
\label{Z6np}\eeq
and
\begin{eqnarray}
{\cal Z}_k \left(\vec a,\E,\EE;q,\bar q,r\right)
&=& \frac{1}{k!}\sum_{\vec{m}\in\mathbb{Z}^{k}} \int_{\mathbb{R}^{k}} \prod_{s=1}^{k} \frac{\mathrm{d} (r\sigma_{s})}{2\pi} e^{-4 \pi i \xi r \sigma_{s}-i \widehat{\theta} m_{s}} 
Z_{\text{gauge}} Z_{IJ}\, Z_{\text{adj}}\label{par}
\end{eqnarray}
with $\widehat{\theta} = \theta + (k-1) \pi$,
\begin{equation}
Z_{\text{gauge}} = \prod_{s<t}^{k}\left(\dfrac{m_{st}^{2}}{4} + r^2\sigma_{st}^{2}\right) \label{gauge}
\end{equation}
and  
\begin{eqnarray}
Z_{IJ} &=& \prod_{s=1}^{k}\prod_{j=1}^{N}\frac{\Gamma\left(-i r \sigma_{s}+i r a_{j} + i r \frac{\epsilon}{2} -\frac{m_{s}}{2}\right)}{\Gamma\left(1+i r \sigma_{s}-i r a_{j} - i r \frac{\epsilon}{2} -\frac{m_{s}}{2}\right)}
\frac{\Gamma\left(i r \sigma_{s}-i r a_{j} + i r \frac{\epsilon}{2} +\frac{m_{s}}{2}\right)}{\Gamma\left(1-i r \sigma_{s}+i r a_{j}- i r \frac{\epsilon}{2} + \frac{m_{s}}{2}\right)} \label{matter}
\\
Z_{\text{adj}} &=& \prod_{s,t=1}^{k}\frac{\Gamma\left(1-i r \sigma_{st}-i r \epsilon-\frac{m_{st}}{2}\right)}{\Gamma\left(i r \sigma_{st}+i r \epsilon-\frac{m_{st}}{2}\right)}
\frac{\Gamma\left(-i r \sigma_{st}+i r \epsilon_{1}-\frac{m_{st}}{2}\right)}{\Gamma\left(1+i r \sigma_{st}-i r \epsilon_{1}-\frac{m_{st}}{2}\right)}
\frac{\Gamma\left(-i r \sigma_{st}+i r \epsilon_{2}-\frac{m_{st}}{2}\right)}{\Gamma\left(1+i r \sigma_{st}-i r \epsilon_{2}-\frac{m_{st}}{2}\right)}\nonumber
\end{eqnarray}

The 1-loop term (\ref{Z61l}) is the perturbative contribution to the partition function in six dimensions, the dependence on the 
radius of the 2-sphere $r$ taking into account the sum over the Kaluza-Klein (KK) modes: it reduces to the 4d perturbative Nekrasov 
partition function in the $r\to 0$ limit when these modes become infinitely massive.
Notice that (\ref{Z61l}) can be written also in the more symmetric form ($\EEE=\frac{i}{r}$)
\beq
{\cal Z}_{1-{\rm loop}}=\prod_{l\neq m}
\Gamma_3\left(a_{lm};\E,\EE,\EEE\right)^{-2} \ \ .
\eeq

The non perturbative term takes into account the contributions of the topological sectors of the gauge theory labeled by 
the second and third Chern character of the gauge bundle, with generating parameters $Q$ and $(q,\bar q)$ respectively.
The six dimensional gauge theory is the effective low energy theory of a system of D5-D1-D(-1) branes on
the minimal resolution of a transversal $A_1$ singularity $\mathbb{C}^2\times T^*S^2 \times \mathbb{C}$,  
where $N$ D5-branes are located on $\mathbb{C}^2\times S^2$, 
$k$ D1 branes are wrapping the two sphere and the D(-1)s are located at the North and the South pole of the sphere, 
the expansion in $(q,\bar q)$ accounting for the two types.
These are nothing but the vortex/anti-vortex contributions of the spherical partition function describing the effective dynamics
of the $k$ D1-branes.
Eq.(\ref{par}) was derived in \cite{Bonelli:2013rja} using the results of \cite{Benini:2012ui,Doroud:2012xw}
applied to the relevant gauged linear sigma model. 
This flows in the IR to the $(2,2)$ supersymmetric non linear sigma model with target space the ADHM instanton moduli space,
the $\Omega$-background being taken into account by the twisted masses $\epsilon_1$ and $\epsilon_2$.

As shown in \cite{Bonelli:2013rja}, eq.(\ref{par}) includes the finite size corrections to the 4d instanton partition 
function due to the KK modes on the two sphere. From a mathematical perspective, it was proposed that these are 
effective world-sheet instantons computing the equivariant Gromov-Witten invariants of the ADHM moduli space.
More precisely, eq.(\ref{par}) can be used to describe the
equivariant quantum cohomology of the ADHM space in terms of a generalization of Givental's ${\cal I}$-function
adapted to non abelian GIT quotients \cite{Bonelli:2013rja,Bonelli:2013mma}. A mathematically rigorous formulation of this generalization has been provided in 
\cite{Ciocan-Maulik}.
The ${\cal I}$-function of the ADHM instanton moduli space can be obtained from a factorized representation of the spherical partition function
\eqref{par} as discussed in detail in \cite{Bonelli:2013rja} and reads
\begin{eqnarray}
{\cal I}_{kN}&=& \sum_{d_1,\ldots , d_k \,\geq \,0} ((-1)^N z)^{d_1+ \ldots +d_k}  \prod_{s=1}^k \prod_{j=1}^N \dfrac{(-r \lambda_{s}-i r a_{j}+ i r \epsilon)_{d_s}}{(1-r \lambda_{s}-i r a_{j})_{d_s}}
\prod_{s<t}^k \dfrac{d_t - d_s - r \lambda_{t} + r \lambda_{s}}{- r \lambda_{t} + r \lambda_{s}}\nonumber\\
&&\dfrac{(1+ r \lambda_{s}- r\lambda_{t}-i r \epsilon)_{d_t - d_s}}{( r \lambda_{s}- r\lambda_{t}+i r \epsilon)_{d_t - d_s}} 
\dfrac{( r \lambda_{s}- r\lambda_{t}+i r \epsilon_{1})_{d_t - d_s}}{(1+ r \lambda_{s}- r\lambda_{t}-i r \epsilon_{1})_{d_t - d_s}}
\dfrac{(r \lambda_{s}- r\lambda_{t}+i r \epsilon_{2})_{d_t - d_s}}{(1+ r \lambda_{s}- r\lambda_{t}-i r \epsilon_{2})_{d_t - d_s}}\nonumber\\
\label{I-inst}\end{eqnarray}
where $\lambda_s$ are the Chern roots of the tautological bundle of the ADHM moduli space.
From the above expression we find that the asymptotic behaviour in $r\to 0$, where $r$ is the radius of the two-sphere, is 
\beq\label{alto}
{\cal I}_{kN}= 1 + r^N {I^{(N)}} +\ldots
\eeq
We recall that the coefficient of the first order term in the small $r$ expansion is identified with the equivariant mirror map. Then from \eqref{alto}
we conclude that  
the equivariant mirror map is trivial, namely ${\cal I}_{k,N}={\cal J}_{k,N}$, for $N>1$, in agreement with the general theorem of \cite{Maulik:2012wi}
on the equivariant quantum cohomology of Nakajima's quiver varieties.

An interesting question to raise is whether a mirror picture 
resumming all the effective world-sheet instantons
can be obtained and what its interpretation from the point of view of integrable systems is.
Answering these questions is the aim of the rest of the paper.

\section{Landau-Ginzburg mirror of the ADHM moduli space and quantum Intermediate Long Wave system}

Let us start by computing the mirror of the ADHM moduli space. This is provided by
a LG model which we study in the Coulomb branch.

A good starting point is to 
define\footnote{We shift $i r \Sigma_s \rightarrow i r \Sigma_s - i r \frac{\epsilon}{2}$ with respect to \cite{Bonelli:2013rja}.} 
\begin{equation}
\Sigma_s = \sigma_s - i\dfrac{m_s}{2 r}
\end{equation}
since this is the twisted chiral superfield corresponding to the superfield strength for the $s$-th vector supermultiplet in the Cartan of $U(k)$. We can now use the procedure described in \cite{Gomis:2012wy}: for every ratio of Gamma functions, we can write 
\begin{equation}
\dfrac{\Gamma(-i r \Sigma)}{\Gamma(1 + i r \overline{\Sigma})} = \int \dfrac{d^2Y}{2\pi}  \text{exp} \Big\{ - e^{-Y} + i r \Sigma Y + e^{-\overline{Y}} + i r \overline{\Sigma} \overline{Y} \Big\} \label{go}
\end{equation}
Here $Y$, $\overline{Y}$ are interpreted as the twisted chiral fields for the matter sector of the mirror Landau-Ginzburg model. Since we want to study the Coulomb branch of this theory in the IR, we have to integrate out the $Y$, $\overline{Y}$ fields. Performing a semiclassical approximation of \eqref{go}, this implies
\begin{equation}
Y = - \ln (- i r \Sigma) \;\;\;\;,\;\;\;\; \overline{Y} = - \ln (i r \overline{\Sigma})
\end{equation}
and we are left with
\begin{equation}
\dfrac{\Gamma(-i r \Sigma)}{\Gamma(1 + i r \overline{\Sigma})} \,\sim \, \text{exp} \Big\{ \omega(- i r \Sigma) - \dfrac{1}{2} \ln (-i r \Sigma) -  \omega(i r \overline{\Sigma}) - \dfrac{1}{2} \ln( i r \overline{\Sigma}) \Big\} \label{exp}
\end{equation}
in terms of the function $\omega(x) = x (\ln x - 1)$.
Defining $t = \xi - i \frac{\theta}{2\pi}$ as the complexified Fayet-Iliopoulos \footnote{The sign of $\theta$ is different from the choice made in \cite{Bonelli:2013rja}.}, equation \eqref{par} becomes
\begin{equation}
\begin{split}
Z_{k,N}^{S^2} &= \frac{1}{k!} \left( \dfrac{ \epsilon}{r \epsilon_1 \epsilon_2 } \right)^{k} \int \prod_{s=1}^{k} \frac{\mathrm{d}^2 (r \Sigma_{s})}{2\pi} \Bigg{\vert}  \left( \dfrac{\prod_{s=1}^{k}\prod_{t \neq s = 1}^{k} D( \Sigma_{st})}{\prod_{s=1}^{k} Q( \Sigma_s)} \right)^{\frac{1}{2}} e^{-\mathcal{W}} \Bigg{\vert}^2 
\end{split}
\end{equation}
where the logarithmic terms in \eqref{exp} (which modify the effective twisted superpotential with respect to the one on $\mathbb{R}^2$) give the measure of the integral in terms of the functions
\begin{equation}
\begin{split}
Q(\Sigma_s) &= r^{2N}\prod_{j=1}^{N}(\Sigma_s - a_j- \frac{\epsilon}{2})(- \Sigma_s + a_j -  \frac{\epsilon}{2})\\
D(\Sigma_{st}) &= \dfrac{( \Sigma_{st})( \Sigma_{st} +  \epsilon)}{( \Sigma_{st} - \epsilon_1)( \Sigma_{st} - \epsilon_2)}\label{factors}
\end{split}
\end{equation}
$\mathcal{W}$ is the effective twisted superpotential of the mirror LG model in the Coulomb branch:
\begin{eqnarray}
\mathcal{W} &=& (2 \pi t - i(k-1)\pi) \sum_{s=1}^k i r \Sigma_s + \sum_{s=1}^k \sum_{j=1}^{N} \left[ \omega(i r \Sigma_s - i r a_j -  i r \frac{\epsilon}{2}) + \omega(- i r \Sigma_s + i r a_j - i r \frac{\epsilon}{2}) \right]\nonumber\\
& + & 
\sum_{\substack{s,t = 1 
}}^k \left[ \omega(i r \Sigma_{st} + i r \epsilon) + \omega(i r \Sigma_{st} - i r \epsilon_1) + \omega(i r \Sigma_{st} - i r \epsilon_2) \right] \label{YY}
\end{eqnarray}
The complex conjugation refers to $\Sigma$ and $t$; in particular, we have
\begin{equation}
\overline{\mathcal{W}(i r \Sigma, t)} = \mathcal{W}(-i r \overline{\Sigma}, \overline{t}) = - \mathcal{W}(i r \overline{\Sigma}, \overline{t})\,\, . \label{vir}
\end{equation}
The function $\mathcal{W}$ coincides with the Yang-Yang function of 
the $gl(N)$ Intermediate Long Wave system 
as proposed in \cite{Litvinov:2013zda}.

Let us now perform a semiclassical analysis around the saddle points of \eqref{YY}.
As we will shortly see, this provides the Bethe-ansatz equations for the quantum integrable system at hand.
By definition, the saddle points are solutions of the equations
\begin{equation}
\dfrac{\partial \mathcal{W}}{\partial (i r \Sigma_s)} = 0
\end{equation}
This implies
\begin{equation}
\begin{split}
& 2\pi t - i (k-1) \pi + \sum_{j=1}^{N} \ln \dfrac{ \Sigma_s -  a_j - \frac{\epsilon}{2}}{- \Sigma_s + a_j -  \frac{\epsilon}{2}} \\
& + \sum_{\substack{t = 1 \\ t\neq s}}^k \ln \dfrac{( \Sigma_{st} + \epsilon)(\Sigma_{st} - \epsilon_1)(\Sigma_{st} - \epsilon_2)}{(- \Sigma_{st} +  \epsilon)(- \Sigma_{st} - \epsilon_1)(- \Sigma_{st} - \epsilon_2)} = 0
\end{split}
\end{equation}
or, by exponentiating and using $(-1)^{k-1} = \prod_{\substack{t = 1 \\ t\neq s}}^k \dfrac{( \Sigma_{st})}{(- \Sigma_{st})}$,
\begin{equation}
\begin{split}
&\prod_{j=1}^{N} (\Sigma_s - a_j - \frac{\epsilon}{2}) \prod_{\substack{t = 1 \\ t\neq s}}^k \dfrac{( \Sigma_{st} - \epsilon_1)(\Sigma_{st} - \epsilon_2)}{(\Sigma_{st})(\Sigma_{st} - \epsilon)}
\\&= e^{-2\pi t}  \prod_{j=1}^{N} (- \Sigma_s + a_j -  \frac{\epsilon}{2}) \prod_{\substack{t = 1 \\ t\neq s}}^k \dfrac{(- \Sigma_{st} - \epsilon_1)(- \Sigma_{st} - \epsilon_2)}{(- \Sigma_{st})(- \Sigma_{st} - \epsilon)} \label{bethe}
\end{split}
\end{equation}
These are the Bethe ansatz equations governing the spectrum of the integrable system for generic $t$ as appeared also in \cite{NOinp,Litvinov:2013zda}. 
To be more precise, remember that $\theta \rightarrow \theta + 2 \pi n$ is a symmetry of the theory; the saddle points will be solutions to 
\begin{equation}
\dfrac{\partial \mathcal{W}}{\partial (i r \Sigma_s)} = 2\pi i n_s
\end{equation}
but this leaves the Bethe ansatz equations \eqref{bethe} unchanged. \\
Around the BO point $t\to\infty$, the solutions to \eqref{bethe} can be labelled by colored partitions of $N$, $\vec{\lambda} = (\lambda_1, \ldots, \lambda_N)$ such that the total number of boxes $\sum_{l=1}^N \vert \lambda_l \vert$ is equal to $k$. In the limit $t \rightarrow \infty$, the roots of the Bethe equations are given by
\begin{equation}
\Sigma_m^{(l)} = a_l + \dfrac{\epsilon}{2} + (i-1) \epsilon_1 + (j-1) \epsilon_2 \;\;\;,\;\;\; m = 1, \ldots , \vert \lambda_l \vert 
\end{equation} 
with $i,j$ running over all possible rows and columns of the tableau $\lambda_l$; those are exactly the poles appearing in the contour integral representation for the 4d Nekrasov partition function \cite{2002hep.th....6161N}. In the large $t$ case, the roots will be given in terms of a series expansion in powers of $e^{-2 \pi t}$.

\subsection{Derivation via large $r$ limit and norm of the ILW wave-functions}

The previous results can also (and maybe better) be understood in terms of a large $r$ limit of \eqref{par}. 
In other words this amounts to set $\epsilon_3\sim 0$ with $\E,\EE$ finite and as such is a six-dimensional analogue of the Nekrasov-Shatashvili limit \cite{Nekrasov:2009rc}. We can use  Stirling's approximation:
\begin{equation}
\begin{split}
\Gamma(z) \,\sim\, \sqrt{2 \pi}\, z^{z-\frac{1}{2}}\,e^{-z} \,(1+ o(z^{-1})) \;\;\;,\;\;\; z \to \infty \\
\Gamma(1+z) \,\sim\, \sqrt{2 \pi}\, z^{z+\frac{1}{2}}\,e^{-z} \,(1+ o(z^{-1})) \;\;\;,\;\;\; z \to \infty 
\end{split}
\end{equation}
which implies
\begin{equation}
\begin{split}
\ln \Gamma(z) \,\sim\, \omega(z) - \dfrac{1}{2} \ln z + \dfrac{1}{2}\ln 2 \pi + o(z^{-1}) \;\;\;,\;\;\; z \to \infty \\
\ln \Gamma(1+z) \,\sim\, \omega(z) + \dfrac{1}{2} \ln z + \dfrac{1}{2}\ln 2 \pi + o(z^{-1}) \;\;\;,\;\;\; z \to \infty 
\end{split}
\end{equation}
Consider for example the contribution from the $I$ field; we have
\begin{equation}
\begin{split}
& \ln \Gamma(-i r \Sigma_s + i r a_j +  i r \frac{\epsilon}{2}) \,\sim\, \omega(-i r \Sigma_s + i r a_j + i r \frac{\epsilon}{2}) - \dfrac{1}{2} \ln (-i r \Sigma_s + i r a_j + i r \frac{\epsilon}{2}) + \dfrac{1}{2}\ln 2 \pi  \\
& \ln \Gamma(1 + i r \overline{\Sigma}_s - i r a_j- i r \frac{\epsilon}{2}) \,\sim\, \omega(i r \overline{\Sigma}_s - i r a_j- i r \frac{\epsilon}{2}) + \dfrac{1}{2} \ln (i r \overline{\Sigma}_s - i r a_j- i r \frac{\epsilon}{2}) + \dfrac{1}{2}\ln 2 \pi  \\
\end{split}
\end{equation}
Doing the limit for all of the fields, we find again 
\begin{equation}
\begin{split}
Z_{k,N}^{S^2} &= \frac{1}{k!} \left( \dfrac{ \epsilon}{r \epsilon_1 \epsilon_2 } \right)^{k} \int \prod_{s=1}^{k} \frac{\mathrm{d}^2 (r \Sigma_{s})}{2\pi} \Bigg{\vert}  \left( \dfrac{\prod_{s=1}^{k}\prod_{t \neq s = 1}^{k} D( \Sigma_{st})}{\prod_{s=1}^{k} Q( \Sigma_s)} \right)^{\frac{1}{2}} e^{-\mathcal{W}} \Bigg{\vert}^2 
\end{split}
\end{equation}
Refining the semiclassical approximation around the saddle points of $\mathcal{W}$ up to quadratic fluctuations, we obtain (eliminating the $k!$ by choosing an order for the saddle points)
\begin{equation}
Z_{k,N}^{S^2} = \Bigg{\vert} e^{-\mathcal{W}_{\text{cr}}} \left( \dfrac{ \epsilon}{r \epsilon_1 \epsilon_2 } \right)^{\frac{k}{2}} \left( \dfrac{\prod_{s=1}^{k}\prod_{t \neq s = 1}^{k} D( \Sigma_{st})}{\prod_{s=1}^{k} Q( \Sigma_s)} \right)^{\frac{1}{2}} \left(\text{Det} \dfrac{\partial^2 \mathcal{W}}{r^2 \partial \Sigma_s \partial \Sigma_t} \right)^{-\frac{1}{2}} \Bigg{\vert}^2  \label{no}
\end{equation}
Apart from the classical term $\vert e^{-\mathcal{W}_{\text{cr}}} \vert^2$, this can be seen as the inverse norm square of the eigenstates of the infinite set of integrals of motion for the ILW system, where each eigenstate corresponds to an $N-$partition $\vec{\lambda}$ and so we can denote it by $\vert \vec{\lambda} \rangle$:
\begin{equation}
Z_{k,N}^{S^2} = \dfrac{\vert e^{-\mathcal{W}_{\text{cr}}} \vert^2}{ \langle \vec{\lambda} \vert \vec{\lambda} \rangle }
\end{equation}
Comparing with \eqref{no}, we find
\begin{equation}
\dfrac{1}{ \langle \vec{\lambda} \vert \vec{\lambda} \rangle } = \Bigg{\vert} \left( \dfrac{ \epsilon}{r \epsilon_1 \epsilon_2 } \right)^{\frac{k}{2}} \left( \dfrac{\prod_{s=1}^{k}\prod_{t \neq s = 1}^{k} D( \Sigma_{st})}{\prod_{s=1}^{k} Q( \Sigma_s)} \right)^{\frac{1}{2}} \left(\text{Det} \dfrac{\partial^2 \mathcal{W}}{r^2 \partial \Sigma_s \partial \Sigma_t} \right)^{-\frac{1}{2}} \Bigg{\vert}^2 
\end{equation}
For real parameters (for example when $t \rightarrow \infty$), this formula agrees with the expression for the norm found in \cite{Litvinov:2013zda}.

\subsection{Quantum ILW Hamiltonians}
\label{sec:Quantum Hamiltonians}

In this subsection we propose that the chiral ring observables of 
the $U(N)$ six-dimensional gauge theory correspond to the set of commuting quantum Hamiltonians of the $gl(N)$ 
ILW system. Due to R-symmetry selection rules, the chiral ring observables vanish in the perturbative sector
and are therefore completely determined by their non-perturbative contributions. These are computed
by the effective two-dimensional GLSM describing D1-branes dynamics in presence of D(-1)s.    
More precisely, chiral observables of the GLSM provide a basis for the quantum 
Hamiltonians of the corresponding integrable system 
\cite{Nekrasov:2009uh, 2009PThPS.177..105N, Nekrasov:2009rc}
. This implies that in our case the quantum
Hamiltonians for the ILW system are given by linear combinations of $\text{Tr}\, \Sigma^n$ operators, for 
generic values of $t$:
\begin{equation}\label{antonio}
\text{ILW quantum Hamiltonians} \;\;\; \longleftrightarrow \;\;\; \text{Tr}\, \Sigma^n (t)
\end{equation}
The calculation of the local chiral ring observables of $U(N)$ gauge theory 
on $\mathbb{C}^2\times S^2$ is analogous to the one on $\mathbb{C}^2$, the crucial difference being 
that in the six dimensional case the bosonic and fermionic zero-modes in the instanton background acquire an extra dependence on the two-sphere coordinates. As a consequence, the sum over the fixed points is replaced by the sum
over the vacua of the effective GLSM giving
\beq
\tr \ e^{\Phi} = \sum_{l=1}^N  \left(e^{a_l}- e^{-\frac{\E+\EE}{2}}(1-e^\E)(1-e^\EE)\sum_m e^{\Sigma_m(t)}\right)
\label{ch}
\eeq 
where $\Sigma_m (t)$ are the solutions of the Bethe equations \eqref{bethe}.
We expect the above formula can be proved in a rigorous mathematical setting in the context of
ADHM moduli sheaves introduced in \cite{emanuel}.
In the $N=2$ case the first few terms read 
\begin{equation}
\begin{split}
& \dfrac{\text{Tr} \Phi^{2}}{2} = a^2 - \E \EE \left( \sum_{m=1}^{\vert \lambda \vert} 1 + \sum_{n=1}^{\vert \mu \vert}  1 \right) \\
& \dfrac{\text{Tr} \Phi^{3}}{3} = -2 \E \EE \left( \sum_{m=1}^{\vert \lambda \vert} \Sigma_m + \sum_{n=1}^{\vert \mu \vert}  \Sigma_n \right) \\
& \dfrac{\text{Tr} \Phi^{4}}{4} = \dfrac{a^4}{2} - 3 \E \EE \left( \sum_{m=1}^{\vert \lambda \vert} \Sigma^2_m  + \sum_{n=1}^{\vert \mu \vert}  \Sigma^2_n \right) - \E \EE\dfrac{\E^2 + \EE^2}{4} \left( \sum_{m=1}^{\vert \lambda \vert} 1  + \sum_{n=1}^{\vert \mu \vert} 1 \right)\\
& \dfrac{\text{Tr} \Phi^{5}}{5} = - 4 \E \EE \left( \sum_{m=1}^{\vert \lambda \vert} \Sigma^3_m  + \sum_{n=1}^{\vert \mu \vert}  \Sigma^3_n \right) - \E \EE(\E^2 + \EE^2) \left( \sum_{m=1}^{\vert \lambda \vert} \Sigma_m  + \sum_{n=1}^{\vert \mu \vert} \Sigma_n \right) \ \ .
\end{split}
\label{nome}
\end{equation}
A check the proposal \eqref{antonio} can be obtained by considering the four dimensional limit where explicit formulae 
are already known. 
Indeed in the four dimensional limit $t\to\pm\infty$ the roots of the Bethe equations reduces to \cite{Litvinov:2013zda} 
\begin{equation}\label{BOBR}
\Sigma_m \;=\; a + \dfrac{\epsilon}{2} + (i-1) \epsilon_1 + (j-1) \epsilon_2 \;=\; a - \dfrac{\epsilon}{2} + i \epsilon_1 + j \epsilon_2 \;\;\;,\;\;\; i,j \geqslant 1 \;\;\;,\;\;\; m = 1, \ldots, \vert \lambda \vert \ \ .
\end{equation}
Consequently, \eqref{ch} reduces to the known formula for the chiral ring observables of four-dimensional $U(N)$ SYM \cite{Losev:2003py,Flume:2004rp}:
\begin{equation}
\begin{split}
\text{Tr} \Phi^{n+1} = \sum_{l=1}^{N} a_l^{n+1} + \sum_{l=1}^{N} \sum_{j=1}^{k_1^{(l)}} \Big[ & \left( a_l + \E \lambda_j^{(l)} + \EE(j-1) \right)^{n+1} - \left( a_l + \E \lambda_j^{(l)} + \EE j \right)^{n+1} \\
& - \left( a_l + \EE(j-1) \right)^{n+1} + \left( a_l + \EE j \right)^{n+1} \Big] \label{tr}
\end{split}
\end{equation}
where $\lambda^{(l)}=\{\lambda^{(l)}_1\ge \lambda^{(l)}_2\ge\ldots\ge\lambda^{(l)}_{k_1^{(l)}}\}, l=1,\dots,N$ indicate colored partitions of the instanton number $k=\sum_{l,j} \lambda_j^{(l)}$.
Since the four-dimensional limit corresponds to the $t\to \infty$ limit, we expect that the above 
chiral observables are related to the quantum Hamiltonians of the BO system. 
For definiteness, let us consider the case $N = 2$.
The higher rank case is discussed in Appendix \ref{BON}. For $N=2$ the Young tableaux correspond to bipartitions ($\lambda$, $\mu$) $= (\lambda_1 \geqslant \lambda_2 \geqslant \ldots, \mu_1 \geqslant \mu_2 \geqslant \ldots)$  such that $\vert \lambda \vert + \vert \mu \vert = k$. 
For Benjamin-Ono, the eigenvalues of the Hamiltonian operators $\mathbf{I}_n$ are given by linear combinations of the eigenvalues of two copies of trigonometric Calogero-Sutherland system 
\cite{Litvinov:2013zda,Alba:2010qc} as  
\begin{equation}
h^{(n)}_{\lambda, \mu} = h^{(n)}_{\lambda}(a) + h^{(n)}_{\mu}(-a)
\end{equation}
with
\begin{equation}\label{h}
h^{(n)}_{\lambda}(a) = \EE \sum_{j=1}^{k_1^{(\lambda)}} \left[ \left( a + \E \lambda_j + \EE \left(j-\frac{1}{2}\right) \right)^n - \left( a + \EE \left(j-\frac{1}{2}\right) \right)^n \right]
\end{equation}
where $k_1^{(\lambda)}$ is the number of boxes in the first row of the partition $\lambda$, and $\lambda_j$ is the number of boxes in the $j$-th column. In particular, $h^{(1)}_{\lambda, \mu} = \E \EE k$.
In terms of \eqref{h}, the $N=2$ chiral observables \eqref{tr} read
\begin{equation}
\begin{split}
\dfrac{\text{Tr} \Phi^{n+1}}{n+1} = 
\dfrac{a^{n+1} + (-a)^{n+1}}{n+1} - \sum_{i=1}^n \dfrac{1+(-1)^{n-i}}{2} \dfrac{n!}{i!(n+1-i)!} \left( \dfrac{\epsilon_2}{2} \right)^{n-i} h^{(i)}_{\lambda, \mu}
\end{split}
\end{equation}
The contributions from $i= 0$, $ i = n+1$ are zero, so they were not considered in the sum. 
The first few cases are:
\begin{equation}\label{few}
\begin{split}
\dfrac{\text{Tr} \Phi^{2}}{2} = a^2 - \E \EE k \hspace{0.5 cm} &, \hspace{0.5 cm}
\dfrac{\text{Tr} \Phi^{3}}{3} = -h^{(2)}_{\lambda, \mu}\\
\dfrac{\text{Tr} \Phi^{4}}{4} = \dfrac{a^4}{2} -h^{(3)}_{\lambda, \mu} - \dfrac{\epsilon_2^2}{4} \E \EE k \hspace{0.5 cm} &, \hspace{0.5 cm}
\dfrac{\text{Tr} \Phi^{5}}{5} = -h^{(4)}_{\lambda, \mu} - \dfrac{\epsilon_2^2}{2} h^{(2)}_{\lambda, \mu}
\end{split}
\end{equation}
We now rewrite the above formulae in terms of the BO Bethe roots \eqref{BOBR}
so that
\begin{equation}
\begin{split}
& h^{(1)}_{\lambda} \;=\; \E \EE \sum_{m=1}^{\vert \lambda \vert} 1 \\
& h^{(2)}_{\lambda} \;=\; 2 \E \EE \sum_{m=1}^{\vert \lambda \vert} \Sigma_m \\
& h^{(3)}_{\lambda} \;=\; 3 \E \EE \sum_{m=1}^{\vert \lambda \vert} \Sigma_m^2 + \E \EE \dfrac{\E^2}{4} \sum_{n=1}^{\vert \lambda \vert} 1 \\
& h^{(4)}_{\lambda} \;=\; 4 \E \EE \sum_{m=1}^{\vert \lambda \vert} \Sigma_m^3 + \E \EE \E^2 \sum_{n=1}^{\vert \lambda \vert} \Sigma_m \, .
\end{split}
\end{equation}

\subsection{Quantum KdV}

Another very interesting limit to analyse is the $\delta\to 0$ limit which provides a connection with quantum KdV
system. Let us recall that KdV is a bi-Hamiltonian system, displaying a further Poisson bracket structure behind
the standard one \eqref{1poisson}, namely 
\beq
\{U(x),U(y)\}=2\left(U(x)+U(y)\right)\delta'(x-y) + \delta'''(x-y)
\label{2poisson}
\eeq
The mapping between the Hamiltonians of the integrable hierarchy with respect to the first and second
Hamiltonian structure can be obtained via the Miura transform 
\beq
U(x) = u_x(x) - u(x)^2
\label{miura}
\eeq
A quantization scheme for KdV system starting from the second Hamiltonian structure was presented in \cite{Bazhanov:1994ft}
where it was shown that the quantum Hamiltonians corresponds to the Casimir operators in the enveloping algebra
${\rm UVir}$.
In particular, the profile function $U(x)$ is the semiclassical limit of the energy-momentum tensor
of the two-dimensional conformal field theory. 

It is interesting to observe that the chiral ring observables of the abelian six-dimensional gauge theory 
provide an alternative quantization of the same system, obtained starting from the first Poisson
bracket structure. Indeed the quantum ILW Hamiltonian $\tr \Phi^3$ reads in the $U(1)$ case
\beq
\begin{split}
H_{ILW}
=&
\left(\E+\EE\right)\sum_{p>0}\frac{p}{2}\frac{q^p+1}{q^p-1}\alpha_{-p}\alpha_p
+
\sum_{p,q >0}\left[\E\EE\alpha_{p+q}\alpha_{-p}\alpha_{-q}-\alpha_{-p-q}\alpha_{p}\alpha_{q}\right] \\
&- \frac{\E+\EE}{2}\frac{q+1}{q-1} \sum_{p>0} \alpha_{-p}\alpha_p 
\label{u1}
\end{split}
\eeq 
where the free field is $\partial\phi = i Q \sum_{k>0} z^k \alpha_k - i Q \E\EE \sum_{k>0} z^{-k}\alpha_{-k}$
and $Q=b+1/b$, $b=\sqrt{\E/\EE}$. This reproduces in the semiclassical limit $b\to 0$ the hydrodynamic profile 
$\partial\phi\to iQ u$ and from \eqref{u1} the ILW Hamiltonian up to and overall factor $-(\E+\EE)$.
Let us notice that due to the twisting with the equivariant canonical bundle of $\mathbb{C}^2$,
the Hermitian conjugation for the oscillators reads $\alpha_k^\dagger = \E\EE \alpha_{-k}$,
$\alpha_{-k}^\dagger = \alpha_{k}/ \E\EE$.
By setting $\theta=0$ and in the $2\pi t =\delta \to 0$ limit \eqref{u1} reduces to
\beq
H_{qKdV}= \delta \left(\E+\EE\right)\sum_{p>0}\frac{(1-p^2)}{12}\alpha_{-p}\alpha_p
+
\sum_{p,q >0}\left[\E\EE\alpha_{p+q}\alpha_{-p}\alpha_{-q}-\alpha_{-p-q}\alpha_{p}\alpha_{q}\right]
\label{qKdV}
\eeq
which in turn corresponds to the quantum KdV Hamiltonian. 
Notice that the extra term in $\tr \Phi^2$ in \eqref{u1}, which is crucial in order to get a finite $t\to 0$ limit,
is the counterpart of the shift in $u_x/\delta$ in the ILW equation \eqref{ilw}.
We expect that the spectrum of the higher quantum KdV Hamiltonians can be obtained by substituting into \eqref{nome} 
the solutions of the $N=1$ Bethe equations \eqref{bethe} expanded around $t=0$.

The alternative expansion in an imaginary dispersion parameter $\theta$ around the dispersionless KdV point $q=1$ of the quantum Hamiltonian has a nice interpretation
in terms of the orbifold quantum cohomology of the symmetric product of points $S^k(\mathbb{C}^2)$.
Indeed when $\delta=0$, namely $q=e^{i\theta}$, the Hamiltonian of the six dimensional abelian gauge theory
can be shown \cite{Bonelli:2013mma} to reduce to that describing the orbifold quantum cohomology of the symmetric product of points
\cite{BG}.

Let us finally remark that also the BLZ quantization scheme can be recovered in the context of gauge theory.
To this end, one has to consider the $U(2)$ case, whose relevant algebra is precisely ${ H}\oplus { Vir}$.
In this case, the $t\to 0$ limit of $gl(2)$ quantum ILW reduces to a decoupled $U(1)$ current and the BLZ system
of quantum Hamiltonians \cite{Litvinov:2013zda}.

\section{Discussion}
\label{sec:conclusion}

The connection between six-dimensional gauge theory and quantum ILW system constitutes the substratum
of the observed equivalences \cite{MNOP,OP,OP2,BP,BG} 
among different enumerative geometry results arising in different limits
of the supersymmetric partition function. More precisely, these can be resumed
in the following Figure 1\footnote{This figure is intentionally similar to the one in \cite{BG}.}.



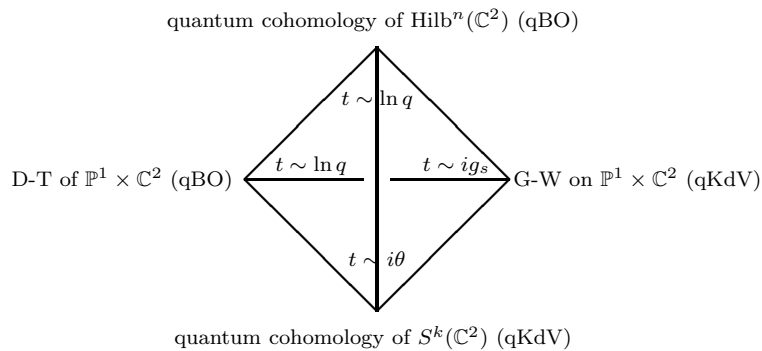
\begin{figure}
\vspace{2cm}
\begin{center}
\scriptsize
\begin{picture}(200,75)(-30,-50)
\thicklines
\put(25,25){\line(1,1){50}}
\put(25,25){\line(1,-1){50}}
\put(125,25){\line(-1,1){50}}
\put(125,25){\line(-1,-1){50}}
\put(75,-25){\line(0,1){100}}
\put(25,25){\line(1,0){45}}
\put(80,25){\line(1,0){45}}
\put(75,85){\makebox(0,0){quantum cohomology of $\text{Hilb}^n(\mathbb{C}^2)$ (qBO) }}
\put(75,55){\makebox(0,0){$t\sim\ln q $}}
\put(75,-35){\makebox(0,0){quantum cohomology of $S^k(\mathbb{C}^2)$ (qKdV) }}
\put(75,-5){\makebox(0,0){$t\sim i\theta$}}
\put(175,25){\makebox(0,0){G-W on $\mathbb{P}^{1}\times \mathbb{C}^{2}$ (qKdV) }}
\put(105,30){\makebox(0,0){$t\sim ig_s$}}
\put(-20,25){\makebox(0,0){D-T of $\mathbb{P}^{1}\times \mathbb{C}^{2}$ (qBO) }}
\put(50,30){\makebox(0,0){$t\sim\ln q$}}
\end{picture}
\end{center}
\caption{The ILW Compass}
\end{figure}


At the North corner the partition function is expanded around $q=e^{-2\pi t}=0$ namely $t\to \infty$
and computes the equivariant quantum cohomology of the Hilbert scheme of points in $\mathbb{C}^2$. This corresponds
to the expansion of the ILW integrable system around the BO point.
At the South corner the expansion is instead around $q=1$ and computes the orbifold quantum cohomology 
of the symmetric product of points in $\mathbb{C}^2$. Here $\delta=0$ and $t\sim i\theta$ where $\theta$ is the expansion parameter for the
orbifold quantum cohomology and the ILW system is expanded around the dispersionless KdV point.   

Equivalent counting problems are represented by the West-East corners.
The West corner provides an alternative interpretation of the expansion around the BO point in terms 
of Donaldson-Thomas invariants of $\mathbb{P}^1\times\mathbb{C}^2$.
Finally the expansion in the East corner corresponds to the all genus Gromov-Witten invariants
of $\mathbb{P}^1\times\mathbb{C}^2$ with genus expansion parameter $g_s\sim - i t$.
 
The above picture is extended by our results also to the non-abelian case. 
The North corner represents the equivariant quantum cohomology of the ADHM instanton moduli space \cite{Bonelli:2013rja} while the West corner gives higher rank Donaldson-Thomas invariants formulated in terms of ADHM moduli sheaves \cite{emanuel}. The South and East corners, while being well defined from the computational viewpoint, still await a rigorous mathematical definition to the best of our knowledge. In particular the South corner should provide the equivariant quantum cohomology of the Uhlenbeck compactification of the instanton moduli space on $S^4$. 
The above Figure 1 indicate dualities of quantum ILW system related to the modular properties of its integral kernel that it would be interesting to analyse.

There is a number of further open questions to be discussed.\\
In this paper integrable systems of hydrodynamical type have been shown to govern BPS states counting in supersymmetric gauge theories.
These are of a different kind and play a much different r\"ole than the ones arising in the Seiberg-Witten theory.
It is of paramount importance to investigate whether an explicit connection between the two can be extabilished.
We observe that while the systems of SW type are related to an effective IR description of the theory, the ones discussed in this paper are 
deeply interconnected with instanton counting and then with the UV degrees of freedom of the gauge theory.
It would be very interesting to see if a connection arises from RG flow arguments.

We have shown that the equations of motion of periodic ILW system arise 
as hydrodynamical limit of the elliptic Calogero-Moser ones. 
On the other hand it is known that the quantum spectrum of the limiting BO$_N$ system
can be described in terms of $N$ copies of the trigonometric Calogero-Sutherland quantum Hamiltonians.
It is thus a pressing question to establish whether a similar relation can be found between
quantum $gl(N)$ ILW and (copies of) quantum elliptic Calogero-Moser.
To this end, and also for other purposes, a generalization of our results to the K-theoretic
setting would be welcome, see \cite{CNO} for a discussion of the abelian case. This passes through the M-theory lift of the geometric set-up, the extra circle
encoding the K-theoretic structure. Doing so, one computes finite $S^2$-size corrections to the
five-dimensional Nekrasov partition function in terms of equivariant quantum K-theory of the ADHM instanton
moduli space, and, from the integrable system viewpoint, provides a connection with Ruijsenaars's relativistic
generalization of Calogero systems. This could also suggest the existence of an integrable 
relativistic generalization of ILW hydrodynamics based on q-deformed Virasoro algebra.

In this paper we pointed out a precise relation between the equivariant quantum cohomology of the
ADHM instanton moduli space and quantum $gl(N)$ ILW. Let us notice that a relation between generalized
two-dimensional topological gravity and classical ILW system has been recently discussed in \cite{Buryak}.
It would be interesting to investigate, for example along the lines of \cite{Brini:2011ij}, whether any relation exists between these results.

The same type of computations presented in this paper can be promptly generalised to other Nakajima quiver varieties whose integrable system description is not known so far.

Finally, we discussed how the four-dimensional limit of our results provides a proof of AGT correspondence
involving $H \oplus W_N$ algebrae in full gauge theoretic terms. Along the same lines one should be able 
to produce a proof also for other classical gauge groups.

\section*{Acknowledgements}
We wish to thank 
N. Nekrasov for a very useful discussion at the early stage of this project, and
M. Bershtein for pointing to our attention \cite{Litvinov:2013zda}.
We thank A. Brini for his careful reading of the manuscript and A.G. Abanov, D.~Diaconescu, B. Dubrovin, A.N.W. Hone, C. Kozcaz, A. Okounkov, V. Roubtsov, R. Santachiara for interesting discussions and comments. 
A.T. thanks the Simons Center for Geometry and Physics for hospitality. 
This research was partly supported by the INFN Research Project GAST ``Nonperturbative dynamics of gauge theory", 
by the INFN Research Project ST\&FI, 
by   PRIN    ``Geometria delle variet\`a algebriche"
and
by  MIUR-PRIN contract 2009-KHZKRX.

\appendix

\section{Details on the proof of \eqref{eq:auxsystem} and \eqref{eq:diffeq}}
\label{eCM}

\subsection{Proof of \eqref{eq:auxsystem}}

First of all we pass to the $\zeta$-function representation of \eqref{eq:auxsystem} by employing the identity
\begin{equation}
\frac{\theta_1^{\prime}\left(\frac{\pi}{L}z\right)}{\theta_1\left(\frac{\pi}{L}z\right)}=\zeta(z)-\frac{2\eta_1}{L}z.
\end{equation}
As was mentioned all the dependence on $\eta_1$ drops out in the result. After doing so and computing $\ddot{x}_j$ from \eqref{eq:auxsystem} we get
\begin{align}
\ddot{x}_j=-G^2\left(L_1+L_2+L_3\right),
\end{align}
where  
\begin{alignat}{3}
 &L_1=&&-\sum_{k=1}^N&&\wp(x_j-y_k)\Big[\sum_{l=1}^N\zeta(x_j-y_l)-\sum_{l\neq j}\zeta(x_j-x_l)+\sum_{l=1}^N\zeta(y_k-x_l)-\sum_{l\neq k}\zeta(y_k-y_l)\Big]  \notag \\ 
& &&+\sum_{k\neq j}&&\wp(x_j-x_k)\Big[\sum_{l=1}^N\zeta(x_j-y_l)-\sum_{l\neq j}\zeta(x_j-x_l)-\sum_{l=1}^N\zeta(x_k-y_l)+\sum_{l\neq k}\zeta(x_k-x_l)\Big] \\
&L_2=&&\frac{2\eta_1}{L}\Bigg\{
&&-\sum_{k\neq j}\left(\wp(x_j-x_k)+\frac{2\eta_1}{L} \right)\Big[\sum_l (x_j-y_l)-\sum_{l\neq j}(x_j-x_l)-\sum_l (x_k-y_l)+\sum_{l\neq k}(x_k-x_l)\Big] \notag \\
& && &&+\sum_k \left(\wp(x_j-y_k)+\frac{2\eta_1}{L} \right)\Big[\sum_l (x_j-y_l)-\sum_{l\neq j}(x_j-x_l)+\sum_l (y_k-x_l)-\sum_{l\neq k}(y_k-y_l)\Big]
\Bigg\}\\
&L_3=&&\frac{2\eta_1}{L}\Bigg\{
&&-\sum_{k}\Big[\sum_l \zeta(x_j-y_l)-\sum_{l\neq j}\zeta(x_j-x_l)+\sum_l\zeta(y_k-x_l)-\sum_{l\neq k}\zeta(y_k-y_l)\Big] \notag \\
& && &&+\sum_{k\neq j}\Big[\sum_l \zeta(x_j-y_l)-\sum_{l\neq j}\zeta(x_j-x_l)-\sum_l\zeta(x_k-y_l)+\sum_{l\neq k}\zeta(x_k-x_l)\Big] 
\Bigg\}
\end{alignat}
The terms $L_2$ and $L_3$ are manifestly vanishing. 
It is slightly more involved to show the vanishing of $L_3$. By collecting sums with common range, 
we have the relation
\begin{equation}
L_3=\frac{2\eta_1}{L}\Bigg\{\Big[\sum_{k\neq j}\big\{\zeta(x_j-x_k)+\sum_{l\neq k}\zeta(x_k-x_l)\big\}\Big]+\Big[(y_j-y_k)\Big]-\Big[(x_j-y_k)\Big]-\Big[(y_j-x_k)\Big]\Bigg\}.
\end{equation}
which vanishes term by term since
\begin{align}
&\sum_{k\neq j}\big\{\zeta(u_j-v_k)+\sum_{l\neq k}\zeta(v_k-u_l)\big\}=\sum_{k\neq j}\big\{\zeta(u_j-v_k)+\zeta(v_k-u_j)+\sum_{l\neq k,j}\zeta(v_k-u_l)\big\} \notag \\
&=\sum_{k\neq j}\sum_{l\neq k,j}\zeta(v_k-u_l)=\sum_{\substack{\text{pairs}(m,n),m\neq n\\(m,n)\neq j}}\bigg[\zeta(v_m-u_n)+\zeta(u_n-v_m)\bigg]=0,
\end{align}
where we used that $\zeta$ is odd. Summarizing, we have $\ddot{x}_j=-G^2 L_1$ which matches 
\eqref{eq:LEoM} in force of the following
identity between Weierstrass $\wp$ and $\zeta$ functions
\begin{align}
\label{eq:rel1}
0&=\sum_{k\neq j}\wp^\prime(x_j-x_k)\notag \\ 
&+\sum_{k=1}^N\wp(x_j-y_k)\Big[\sum_{l=1}^N\zeta(x_j-y_l)-\sum_{l\neq j}\zeta(x_j-x_l)+\sum_{l=1}^N\zeta(y_k-x_l)-\sum_{l\neq k}\zeta(y_k-y_l)\Big] \notag \\ 
&-\sum_{k\neq j}\wp(x_j-x_k)\Big[\sum_{l=1}^N\zeta(x_j-y_l)-\sum_{l\neq j}\zeta(x_j-x_l)-\sum_{l=1}^N\zeta(x_k-y_l)+\sum_{l\neq k}\zeta(x_k-x_l)\Big].
\end{align}
We prove this identity using Liouville's theorem. Let us denote the right hand side by \newline
$R\left(x_j;\{x_k\}_{k\neq j},\{y_k\}_{k=1}^N\right)$. $R$ is a symmetric function under independent permutations of $\{x_k\}_{k\neq j}$ and $\{y_k\}_{k=1}^N$, respectively. 
Next, we show double periodicity in all variables. Although the $\zeta$'s introduce shifts, these cancel each other\footnote{All $\zeta$'s appear in pairs, where a given variable appears with positive and negative signs in the argument.}, so double periodicity follows immediately. 
The non-trivial step is to show holomorphicity. First, the relation should hold for all $j$. In particular we can choose $j=1$, other cases are obtained just by relabeling. By double periodicity we can focus only on poles at the origin, so there will be poles in $x_j-y_k$ and $x_j-x_l$, $l\neq j$. By the symmetries described above we have to check only three cases: $x_1-y_1$, $x_2-y_1$ and $x_1-x_2$. To do so, we use the Laurent series for $\wp$ and $\zeta$
\begin{alignat}{2}\label{eq:Lseries}
\wp(z)&=\frac{1}{z^2}+\wp^R(z),\;\;&&\wp^R(z)=\sum_{n=1}^\infty c_{n+1}z^{2n} \notag \\
\zeta(z)&=\frac{1}{z}+\zeta^R(z),&&\zeta^R(z)=-\sum_{n=1}^\infty \frac{c_{n+1}}{2n+1}z^{2n+1}
\end{alignat}
Let us now show the vanishing of the residues at each pole.
\subsubsection*{Pole in $x_2-y_1$}
There are only two terms in \eqref{eq:rel1} contributing
\begin{align}
&\zeta(x_2-y_1)\Big[\wp(x_1-x_2)-\wp(x_1-y_1)\Big] \notag \\
&\sim\frac{1}{x_2-y_1}\Big[\frac{1}{(x_1-x_2)^2}-\frac{1}{(x_1-y_1)^2}+\sum_{n\geq 1}c_{n+1}\left((x_1-x_2)^{2n}-(x_1-y_1)^{2n}\right)\Big] \notag \\
&=\frac{x_2-y_1}{x_2-y_1}\Big[\frac{1}{(x_1-x_2)^2(x_1-y_1)}+\sum_{n\geq 1}c_{n+1}\sum_{k=1}^{2n}\binom{2n}{k}(-1)^kx_1^{2n-k}\sum_{l=0}^{k-1}x_2^{k-1-l}y_1^l\Big].
\end{align}
So indeed the residue vanishes.
\subsubsection*{Pole in $x_1-y_1$}
The terms contributing to this pole read
\begin{alignat}{2}
&\wp(x_1-y_1)\sum_{k\neq 1}\bigg\{\big[\zeta(&&x_1-y_k)-\zeta(y_1-y_k)\big]-\big[\zeta(x_1-x_k)-\zeta(y_1-x_k)\big]\bigg\} \notag \\
&+\zeta(x_1-y_1)\sum_{k\neq 1}\Big[&&\wp(x_1-y_k)-\wp(x_1-x_k)\Big] \notag \\
&\sim\frac{1}{(x_1-y_1)^2}\sum_{k\neq 1}
\bigg\{
&&\Big[\frac{1}{x_1-y_k}-\frac{1}{y_1-y_k}\Big]-\Big[\frac{1}{x_1-x_k}-\frac{1}{y_1-x_k}\Big] \notag \\
& &&+\Big[\zeta^R(x_1-y_k)-\zeta^R(y_1-y_k)\Big]-\Big[\zeta^R(x_1-x_k)-\zeta^R(y_1-x_k)\Big]
\bigg\} \notag \\
&+\frac{1}{x_1-y_1}\sum_{k\neq 1}\Big[\wp^R(&&x_1-y_k)-\wp^R(x_1-x_k)+\frac{1}{(x_1-y_k)^2}-\frac{1}{(x_1-x_k)^2}\Big].
\end{alignat}
Collecting all the rational terms gives a regular term
\begin{equation}
\sum_{k\neq 1}\Big[\frac{1}{(x_1-x_k)^2(y_1-x_k)}-\frac{1}{(x_1-y_k)^2(y_1-y_k)}\Big]
\end{equation}
and we stay with the rest
\begin{align}
\sum_{k\neq 1}\frac{1}{x_1-y_1}\Bigg\{\wp^R(x_1-y_k)-\wp^R(x_1-x_k)+\frac{1}{x_1-y_1}&\Big[\left(\zeta^R(x_1-y_k)-\zeta^R(y_1-y_k)\right) \notag \\
&-\left(\zeta^R(x_1-x_k)-\zeta^R(y_1-x_k)\right)\Big] \Bigg\}.
\end{align}
In the following we show that the terms in the square parenthesis in the above formula
factorizes a term $(x_1-y_1)$ which, after combining with the rest, cancels the pole completely. Indeed, 
we just use \eqref{eq:Lseries} and binomial theorem to get
\begin{align}
&\Big[\ldots\Big]=-(x_1-y_1)\sum_{n\geq 1}\frac{c_{n+1}}{2n+1}\sum_{l=1}^{2n}\binom{2n+1}{l}(-1)^l\left(y_k^{2n+1-l}-x_k^{2n+1-l}\right)\sum_{m=0}^{l-1}y_1^{l-1-m}x_1^m \notag \\
&\wp^R(x_1-y_k)-\wp^R(x_1-x_k)=\sum_{n\geq 1}c_{n+1}\sum_{l=1}^{2n}\binom{2n}{l-1}(-1)^lx_1^{l-1}\left(y_k^{2n+1-l}-x_k^{2n+1-l}\right)
\end{align}
and after combining these two terms we get
\begin{align}\label{noi}
\Big\{\ldots\Big\}=\sum_{n\geq 1}c_{n+1}\sum_{l=1}^{2n}\binom{2n}{l-1}(-1)^l\left(y_k^{2n+1-l}-x_k^{2n+1-l}\right)\Big[x_1^{l-1}-\frac{1}{l}\sum_{m=0}^{l-1}y_1^{l-1-m}x_1^m\Big],
\end{align}
however the terms in the square brackets of \eqref{noi} factorizes once more a term $(x_1-y_1)$
\begin{align}
\Big[\ldots\Big]=(x_1-y_1)\frac{1}{l}\sum_{m=1}^{l-1}(l-m)x_1^{l-1-m}y_1^{m-1}
\end{align}
so that we end up with a regular term
\begin{align}
\sum_{k\neq 1}\sum_{n\geq 1}c_{n+1}\sum_{l=1}^{2n}\binom{2n}{l-1}\frac{(-1)^l}{l}\left(y_k^{2n+1-l}-x_k^{2n+1-l}\right)\sum_{m=1}^{l-1}(l-m)x_1^{l-1-m}y_1^{m-1}.
\end{align}
Summarizing, we have shown the vanishing of the residue at the pole in $(x_1-y_1)$ 
and we now move on to the last one.
\subsubsection*{Pole in $x_1-x_2$}
Analysis of \eqref{eq:rel1} gives the following terms contributing to this pole
\begin{align}
&\wp^\prime(x_1-x_2)+\zeta(x_1-x_2)\Big[\sum_{k\neq 1,2}\wp(x_1-x_k)-\sum_k\wp(x_1-y_k)\Big] \notag \\
&-\wp(x_1-x_2)\Big[\sum_k\zeta(x_1-y_k)-\sum_{k\neq 1}\zeta(x_1-x_k)-\sum_k\zeta(x2-y_k)+\sum_{k\neq 2}\zeta(x_2-x_k)\Big].
\end{align}
In analogy with the previous case let us first deal with the rational terms
\begin{align}
&\frac{-2}{(x_1-x_2)^3}+\frac{1}{x_1-x_2}\Big[\sum_{k\neq 1,2}\frac{1}{(x_1-x_k)^2}-\sum_k\frac{1}{(x_1-y_k)^2}\Big] \notag \\
&-\frac{1}{(x_1-x_2)^2}\Big[\frac{-2}{x_1-x_2}+\sum_k\left(\frac{1}{x_1-y_k}-\frac{1}{x_2-y_k}\right)-\sum_{k\neq 1,2}\left(\frac{1}{x_1-x_k}-\frac{1}{x_2-x_k}\right)\Big] \notag \\
&=\sum_k\frac{1}{(x_1-y_k)^2(x_2-y_k)}-\sum_{k\neq 1,2}\frac{1}{(x_1-x_k)^2(x_2-x_k)},
\end{align}
which give a regular contribution as we wanted. For the remaining terms we can write, 
using the same methods as above
\begin{alignat}{2}
&\frac{1}{x_1-x_2}\Bigg\{\sum_{k\neq 1,2}\wp^R(x_1-x_k)-\sum_k\wp^R(x_1-y_k)-\frac{1}{x_1-x_2}&&\Big[\sum_k\left(\zeta(x_1-y_k)-\zeta(x_2-y_k)\right) \notag \\
& &&-\sum_{k\neq 1,2}\left(\zeta(x_1-x_k)-\zeta(x_2-x_k)\right)\Big]\Bigg\} \notag \\
&=\sum_{n\geq 1}c_{n+1}\sum_{l=1}^{2n+1}\binom{2n}{l-1}\frac{(-1)^l}{l}\sum_{m=1}^{l-1}(l-m)x_1^{l-1-m}x_2^{m-1}\Big[&&\sum_{k\neq 1,2}x_k^{2n+1-l}-\sum_ky_k^{2n+1-l}\Big],
\end{alignat}
which explicitly shows the vanishing of the residue of this last pole.
 
We just showed that $R\left(x_j;\{x_k\}_{k\neq j},\{y_k\}_{k=1}^N\right)$ is holomorphic in the whole complex plane for all variables. Liouville's theorem then implies it must be a constant. Hence we can set any convenient values for the variables to show this constant to be zero. Taking the limit $y_k\to 0$ for all $k$ we get
\begin{align}
&-\lim_{y_k\to 0}\sum_k\wp(x_1-y_k)\sum_{l\neq k}\frac{1}{y_k-y_l}+\sum_{k\neq 1}\wp^\prime(x_1-x_k) 
+N\wp(x_1)\Big[N\zeta(x_1)-\sum_{k\neq 1}\zeta(x_1-x_k)-\sum_k\zeta(x_k)\Big] \notag \\
&-\sum_{k\neq 1}\wp(x_1-x_k)\Big[N\zeta(x_1)-\sum_{l\neq 1}\zeta(x_1-x_l)-N\zeta(x_k)+\sum_{l\neq k}\zeta(x_k-x_l)\Big]
\end{align}
The first term can be written as
\begin{align}
\lim_{y_k\to 0}\sum_{\substack{\text{pairs}(m,n),m\neq n\\m,n\in\{1,\ldots,N\}}}\frac{1}{y_n-y_m}\Big[\wp^\prime(x_1)(y_n-y_m)+\mathcal{O}\left((y_n-y_m)^2\right)\Big]=\frac{N(N-1)}{2}\wp^\prime(x_1)
\end{align}
Sending $x_k\to 0$, $k\neq 1$ simplifies $R$ further
\begin{align}
&\left(N-1\right)\left(\frac{N}{2}+1\right)\wp^\prime(x_1)-(N-1)\wp(x_1)\zeta(x_1) \notag \\
&+\lim_{\substack{x_k\to 0\\k\neq 1}}\bigg\{\sum_{k\neq 1}\wp(x_1-x_k)\Big[N\zeta(x_k)-\sum_{l\neq k}\zeta(x_k-x_l)\Big]-N\wp(x_1)\sum_{k\neq 1}\zeta(x_k)\bigg\},
\end{align}
where the second line yields
\[
\lim_{\substack{x_k\to 0\\k\neq 1}}\bigg\{\underbrace{N\sum_{k\neq 1}\frac{1}{x_k}\Big[\wp(x_1-x_k)-\wp(x_1)\Big]}_{-N(N-1)\wp^\prime(x_1)}\underbrace{-\sum_{k\neq 1}\wp(x_1-x_k)\sum_{l\neq k}\zeta(x_k-x_l)}_{(N-1)\wp(x_1)\zeta(x_1)+\frac{(N-1)(N-2)}{2}\wp^\prime(x_1)}\bigg\}.
\]
Putting everything together we finally obtain
\[
 \text{const}=\lim_{\substack{y_k\to 0\\x_l\to 0,l\neq 1}}R(\ldots)=0 \;\Longrightarrow\;R(\ldots)=0,
\]
which concludes the proof of \eqref{eq:rel1}.

\subsection{Proof of \eqref{eq:diffeq}}

By simplifying the left hand side of \eqref{eq:diffeq} one gets
\begin{align}
\sum_{j=1}^N\Bigg\{&G\Big[\wp(z-x_j)\zeta(z-x_j)+\frac{1}{2}\wp^\prime(z-x_j)\Big]+G\Big[\wp(z-y_j)\zeta(z-y_j)+\frac{1}{2}\wp^\prime(z-y_j)\Big] \notag \\
&+\wp(z-x_j)\Big[-i\dot{x}_j-G\sum_{k=1}^N\zeta(z-y_k)+G\sum_{k\neq j}\zeta(z-x_k)\Big] \notag \\
&+\wp(z-y_j)\Big[i\dot{y}_j-G\sum_{k=1}^N\zeta(z-x_k)+G\sum_{k\neq j}\zeta(z-y_k)\Big] \notag \\
&+G\frac{2\eta_1}{L}\Big[i\dot{y}_j-i\dot{x}_j+G\left(\wp(z-y_j)-\wp(z-x_j)\right)\sum_k (y_k-x_k)\Big]
\Bigg\}.
\end{align}
Going on-shell w.r.t. auxiliary system \eqref{eq:auxsystem}, we arrive at
\begin{equation}
\text{LHS}=X_1+X_2,
\end{equation}
where
\begin{alignat}{2}
&X_1=\sum_{j=1}^N&&\Bigg\{\frac{1}{2}\wp^\prime(z-x_j)+\wp(z-x_j)\Big[\sum_{k=1}^N\left(\zeta(z-x_k)-\zeta(z-y_k)+\zeta(x_j-y_k)\right)-\sum_{k\neq j}\zeta(x_j-x_k)\Big] \notag \\
& &&+\frac{1}{2}\wp^\prime(z-y_j)+\wp(z-y_j)\Big[\sum_{k=1}^N\left(\zeta(z-y_k)-\zeta(z-x_k)+\zeta(y_j-x_k)\right)-\sum_{k\neq j}\zeta(y_j-y_k)\Big]
\Bigg\} \notag \\
&X_2=G^2&&\frac{2\eta_1}{L}\sum_{j=1}^N\sum_{k\neq j}\Big\{
\zeta(y_j-x_k)+\zeta(x_j-y_k)-\zeta(y_j-y_k)-\zeta(x_j-x_k)
\Big\}.
\end{alignat}
It is easy to see that $X_2$ vanishes, since we can rearrange the sum to pairs of $\zeta$'s with positive and negative arguments respectively
\begin{alignat}{2}
X_2&=G^2\frac{2\eta_1}{L}\sum_{\substack{\text{pairs}(m,n),m\neq n\\m,n\in\{1,\ldots,N\}}}&&\Bigg\{\bigg[\zeta(y_m-x_n)+\zeta(x_n-y_m)\bigg]+\bigg[\zeta(x_m-y_n)+\zeta(y_n-x_m)\bigg] \notag \\
& &&-\bigg[\zeta(x_m-x_n)+\zeta(x_n-x_m)\bigg]-\bigg[\zeta(y_m-y_n)+\zeta(y_n-y_m)\bigg]
\Bigg\} \notag \\
&=0.
\end{alignat}
The vanishing of $X_1$ looks more intriguing, but actually reduces to the already proven relation \eqref{eq:rel1}. Indeed, we can write $X_1$ as
\[
 X_1=\frac{1}{2(N-1)}\sum_{j=1}^{N}\Big[R\left(\{x\},\{y\}\right)\Bigr\rvert_{x_j=z}+R\left(\{x\}\leftrightarrow\{y\}\right)\Bigr\rvert_{y_j=z}\Big]=0,
\]
which concludes the proof of \eqref{eq:diffeq}.

\section{Fock space formalism for the equivariant quantum cohomology of the ADHM moduli space}
\label{NB}

Let us recall the Fock space description of the equivariant cohomology of the Hilbert scheme of points of $\mathbb{C}^2$ -- introduced in
\cite{Grojo,NakaBook} -- following the notation of \cite{BG} and \cite{OP}.
One introduces creation-annihilation operators $\alpha_k$, $k\in \mathbb{Z}$
obeying the Heisenberg algebra
\beq
[\alpha_p,\alpha_q] = p\delta_{p+q}
\label{o}
\eeq
Positive modes annihilate the vacuum 
\beq
\alpha_p |\emptyset\rangle = 0 \ \ , p>0
\eeq
and the natural basis of the Fock space is given by
\beq
|Y\rangle = \frac{1}{|Aut(Y)|\prod_i Y_i}\prod_i \alpha_{Y_i}|\emptyset\rangle
\label{basis}
\eeq
where $|Aut(Y)|$ is the order of the automorphism group of the partition and $Y_i$
are the lengths of the columns of the Young tableau $Y$.
The number of boxes of the Young tableau is counted by  
the eigenvalue of the energy operator $K=\sum_{p>0}\alpha_{-p}\alpha_p$.
Fix now the subspace ${\rm Ker}(K-k)$ with $k\in\mathbb{Z}_+$
and allow linear combinations with coefficients being rational functions of the 
equivariant weights. This space is identified with the equivariant cohomology
$H^*_T\left({\cal M}_{k,1},\mathbb{Q}\right)$.
Explicitly
\beq
|Y\rangle\in H^{2n-2\ell(Y)}_T\left({\cal M}_{k,1},\mathbb{Q}\right),
\eeq
where $\ell(Y)$ denotes the number of parts of the partition $Y$.

According to \cite{OP}, the generator of the small quantum cohomology is given by the state
$|D\rangle=-|2,1^{k-2}\rangle$ describing the divisor which corresponds to 
the collision of two point-like instantons.

The operator generating the quantum product by $|D\rangle$ is given by
the quantum Hamiltonian
\beq
H_D
\equiv
\left(\E+\EE\right)\sum_{p>0}\frac{p}{2}\frac{(-q)^p+1}{(-q)^p-1}\alpha_{-p}\alpha_p
+
\sum_{p,q >0}\left[\E\EE\alpha_{p+q}\alpha_{-p}\alpha_{-q}-\alpha_{-p-q}\alpha_{p}\alpha_{q}\right]
-
\frac{\E+\EE}{2}\frac{(-q)+1}{(-q)-1} K
\label{ham}
\eeq
which can be recognized as the fundamental quantum Hamiltonian of the ILW system.
The generalization of the Fock space formalism to the rank $N$ ADHM instanton moduli space was given by Baranovsky in \cite{Bara}
in terms of $N$ copies of Nakajima operators as $\beta_k = \sum_{i=1}^N \alpha_k^{(i)}$
. For example, in the $N=2$ case the quantum Hamiltonian becomes (modulo terms proportional to the quantum momentum) \cite{Maulik:2012wi}
\begin{equation}\label{MO}
\begin{split}
H_D=& \dfrac{1}{2} \sum_{i=1}^2 \sum_{n,k>0} [\E\EE \alpha_{-n}^{(i)}\alpha_{-k}^{(i)}\alpha_{n+k}^{(i)} - \alpha_{-n-k}^{(i)}\alpha_{n}^{(i)}\alpha_{k}^{(i)}] \\
&- \dfrac{\E + \EE}{2} \sum_{k>0} k[ \alpha_{-k}^{(1)}\alpha_{k}^{(1)} + \alpha_{-k}^{(2)}\alpha_{k}^{(2)} + 2 \alpha_{-k}^{(2)}\alpha_{k}^{(1)}] \\
& -(\E + \EE)\sum_{k>0} k \dfrac{q^{k}}{1-q^{k}} [ \alpha_{-k}^{(1)}\alpha_{k}^{(1)} + \alpha_{-k}^{(2)}\alpha_{k}^{(2)} + \alpha_{-k}^{(2)}\alpha_{k}^{(1)} + \alpha_{-k}^{(1)}\alpha_{k}^{(2)}]
\end{split}
\end{equation}
This is the same as the $I_3$ Hamiltonian for $gl(2)$ ILW given in \cite{Litvinov:2013zda}: 
\begin{equation}
\begin{split}
I_3 = \sum_{k\neq 0} L_{-k}a_k + 2 i Q \sum_{k>0} k a_{-k}a_k \dfrac{1+q^{k}}{1-q^{k}} + \dfrac{1}{3} \sum_{n+m+k=0} a_n a_m a_k 
\end{split}
\end{equation}
In fact, after rewriting the Virasoro generators in terms of Heisenberg generators according to
\begin{equation}
L_n = \sum_{k \neq \{0, n\}} c_{n-k} c_k + i (n Q - 2 P)c_n \;\;\;,\;\;\; [c_m,c_n] = \frac{m}{2}\delta_{m+n,0}
\end{equation}
and ignoring terms proportional to the momentum, we arrive to
\begin{equation}
\begin{split}
I_3 =& \sum_{n,k>0} [a_{-n-k} c_n c_k + 2 a_{-n} c_{-k }c_{n+k} + 2 c_{-n-k} c_n a_k + c_{-n} c_{-k }a_{n+k} ] 
\\
& 
+ 2 i Q \sum_{k> 0} k[a_{-k} a_k - \dfrac{1}{2} (c_{-k}a_k - a_{-k}c_k)]\\
&+ 4 i Q \sum_{k>0} k a_{-k}a_k \dfrac{q^{k}}{1-q^{k}} + \sum_{n,k>0} a_{-n-k} a_n a_k + \sum_{n,k>0} a_{-n} a_{-k }a_{n+k} 
\end{split}
\end{equation}
where we used
\begin{equation}
\sum_{k\neq 0} \sum_{n \neq \{0, -k\}} c_{-n-k} c_n a_k = \sum_{n,k>0} [a_{-n-k} c_n c_k + 2 a_{-n} c_{-k }c_{n+k} + 2 c_{-n-k} c_n a_k + c_{-n} c_{-k }a_{n+k} ] 
\end{equation}
The $a_k$ can be related with the Baranovsky operators. Finally, by making the substitution
\begin{equation}
a_k = -\dfrac{i}{\sqrt{\E\EE}}\dfrac{\alpha_k^{(1)} + \alpha_k^{(2)}}{2} \;\;\;,\;\;\; c_k = - \dfrac{i}{\sqrt{\E\EE}}\dfrac{\alpha_k^{(1)} - \alpha_k^{(2)}}{2}
\end{equation}
for positive modes and 
\begin{equation}
a_{-k} = i \sqrt{\E\EE} \ \dfrac{\alpha_{-k}^{(1)} + \alpha_{-k}^{(2)}}{2} \;\;\;,\;\;\; c_{-k} = i \sqrt{\E\EE}\ \dfrac{\alpha_{-k}^{(1)} - \alpha_{-k}^{(2)}}{2}
\end{equation}
for the negative ones,
we obtain
\begin{equation}
\begin{split}
I_3 =& \dfrac{i}{2\sqrt{\E\EE}} \sum_{n,k>0} [\E\EE \alpha_{-n}^{(1)} \alpha_{-k }^{(1)}\alpha_{n+k}^{(1)} - \alpha_{-n-k}^{(1)} \alpha_n^{(1)} \alpha_k^{(1)}
 + \E\EE \alpha_{-n}^{(2)} \alpha_{-k }^{(2)}\alpha_{n+k}^{(2)} - \alpha_{-n-k}^{(2)} \alpha_n^{(2)} \alpha_k^{(2)} ]\\
& + \dfrac{ i Q}{2} \sum_{k>0} k [\alpha_{-k }^{(1)}\alpha_{k }^{(1)} + \alpha_{-k }^{(2)}\alpha_{k }^{(2)} + 2 \alpha_{-k }^{(2)}\alpha_{k }^{(1)}] 
\\
& + iQ \sum_{k>0} k \dfrac{q^{k}}{1-q^{k}} [\alpha_{-k }^{(1)}\alpha_{k }^{(1)} + \alpha_{-k }^{(2)}\alpha_{k }^{(2)} + \alpha_{-k }^{(1)}\alpha_{k }^{(2)} + \alpha_{-k }^{(2)}\alpha_{k }^{(1)}]
\end{split}
\end{equation}
in agreement with \eqref{MO}. 

\section{BO$_N$ Hamiltonians}
\label{BON}

In Section \ref{sec:Quantum Hamiltonians} we observed that the spectrum of the chiral operator 
$\text{Tr} \Phi^{n+1}$ can be expressed as a linear combination of the eigenvalues of the integrals of motion 
(IMs) of the Benjamin-Ono integrable system\footnote{This is was checked up to $n=4$, where explicit results 
for the eigenvalues of the IMs are available.}. 
We showed explicitly the connection between $SU(2)$ $\mathcal{N} = 2$ supersymmetric Yang-Mills theory and $\text{Vir}\oplus\text{H}$ CFT. 
In this Appendix we consider the $SU(N)$ gauge theory versus $\text{W}_{N}\oplus\text{H}$ algebra, 
focusing mainly on $I_3$, which we identify as the basic Hamiltonian, whose spectrum was computed in \cite{Estienne:2011qk}. As a preliminary check and also to build the dictionary between \cite{Estienne:2011qk} and \cite{Litvinov:2013zda} we can specialize to the $\text{Vir}\oplus\text{H}$ case\footnote{In \cite{Estienne:2011qk} the eigenvalues were computed for a special class of eigenstates. In general, the eigenvalues depend on the momentum $P$, which characterizes the eigenstates, i.e. does not enter into the IMs. So picking a special class of eigenstates translates into setting a given value of the momentum $P=P_{*}$.}. The dictionary is obtained by direct comparison of explicit expressions for IMs and their eigenvalues and can be found in Table \ref{tab:SantaLitdict}.
\begin{savenotes}
\begin{table}
\centering
\begin{tabular}{c|c|c}
 Litvinov & Estienne \textit{et al.} & gauge theory 
\\ \hline
   $b$    &  	$i\sqrt{g}$          & $\epsilon_2$ \footnote{Here we are taking $\epsilon_1\epsilon_2=1$.} \\ \hline
  $a_k$   &   $\sqrt{2}a_k$	     &
\\ \hline
  $P_*$   & special eigenstates      &    $a_*$
\\ \hline   
$b\left(h_{\lambda}^{(2)}(P)-2P |\lambda|\right)$ & $e_{\lambda}^{(3),+}(g)$
\end{tabular}
\caption{Dictionary between \cite{Estienne:2011qk} and \cite{Litvinov:2013zda}}
\label{tab:SantaLitdict}
\end{table}
\end{savenotes}

Comparing the expressions for $I_{3}^{+}(g)$ in \cite{Estienne:2011qk} and $I_2$ in \cite{Litvinov:2013zda} (the labelling is unfortunately shifted) we get
\begin{align}
\label{eq:SantaLitenergy}
I_{3}^{+}(g)=2ibI_2 \;\;\Longrightarrow\;\; 
E_{\overrightarrow{\lambda}}^{(3),+}(g)&=2ib\left(-\frac{i}{2}\left.h_{\overrightarrow{\lambda}}^{(2)}\right|_{P=P_{*}}\right)\notag \\
&=b\left.h_{\overrightarrow{\lambda}}^{(2)}\right|_{P=P_{*}}
\end{align}
To highlight how one picks the special value $P_*$ let us still concentrate only on the $\text{Vir}\oplus\text{H}$ case. Taking the result for $E_{\overrightarrow{\lambda}}^{(3),+}(g)$ from \cite{Estienne:2011qk} and using the third row of table \ref{tab:SantaLitdict} we can write
\begin{align}
 E_{(\lambda,\mu)}^{(3),+}(g)&=e_{\lambda}^{(3),+}(g)+e_{\mu}^{(3),+}(g)-\sqrt{2g}(q-\alpha_0)(|\lambda|-|\mu|) \notag\\
&=bh_{(\lambda,\mu)}^{(2)}(P)+b\left[\sqrt{2}i(q-\alpha_0)-2P\right](|\lambda|-|\mu|),
\end{align}
where $\alpha_0=\frac{i}{\sqrt{2}}Q$ and $q$ is a charge for the zero mode $b_0$ of an auxiliary bosonic field, $b_0|q\rangle=q|q\rangle$. By imposing \eqref{eq:SantaLitenergy} the bracket $\left[\ldots\right]$ is forced to vanish, which leads to
\begin{equation}
P_*=\frac{i}{\sqrt{2}}(q-\alpha_0)
\end{equation}
Finally, concluding the $\text{Vir}\oplus\text{H}$ CFT or $SU(2)$ gauge theory respectively, we get for $\overrightarrow{\lambda}=(\lambda,\mu)$
\begin{equation}
E_{\overrightarrow{\lambda}}^{(3),+}(g)=b\left.h_{\overrightarrow{\lambda}}^{(2)}\right|_{P=P_{*}}=-\epsilon_2\left.\frac{\text{Tr}\Phi_{\overrightarrow{\lambda}}^{3}}{3}\right|_{a=a_*}
\end{equation}

At this point we are ready to make connection between the $\text{W}_{N-1}\oplus\text{H}$ CFT and $SU(N)$ gauge theory for $I_3^+(g)$ and $\text{Tr}\Phi^3$. First, we write the result for $  E_{\overrightarrow{\lambda}}^{(3),+}(g)$ \cite{Estienne:2011qk} and manipulate it to a more convenient form for us
\begin{alignat}{2}
\label{eq:energy}
&E_{\overrightarrow{\lambda}}^{(3),+}(g)&=&\sum_{l=1}^{N}e_{\lambda_l}^{(3),+}+(1-g)\sum_{l=1}^{N}(N+1-2l)|\lambda_l| \notag \\
&{}&=&\epsilon_2^2\sum_{l=1}^N\sum_{j=1}^{\#\text{rows}(\lambda_l)}\Bigg\{\left(a_l+\epsilon_1|\text{row}_j(\lambda_l)|+\epsilon_2\left(j-\frac{1}{2}\right)\right)^2-\left(a_l+\epsilon_2\left(j-\frac{1}{2}\right)\right)^2\Bigg\} \notag\\
&{}&{}&-2\epsilon_2\sum_{l=1}^{N}a_l\lambda_l+(1+\epsilon_2^2)\sum_{l=1}^{N}(N+1-2l)|\lambda_l|.
\end{alignat}
Then we need also to rewrite the expression for $\text{Tr}\Phi^{n+1}$ \eqref{tr}
\begin{align}
\label{eq:Trphi}
\text{Tr}\Phi_{\overrightarrow{\lambda}}^{n+1}=\sum_{l=1}^{N}a_l^{n+1}+&\sum_{l=1}^{N}\sum_{j=1}^{\#\text{rows}(\lambda_l)}(-\epsilon_2)\sum_{i=1}^{n}{n+1\choose{i}}\left(\frac{\epsilon_2}{2}\right)^{n-i}\frac{1+(-1)^{n-i}}{2} \notag \\
&\times\Bigg[\left(a_l+\epsilon_1|\text{row}_j(\lambda_l)|+\epsilon_2\left(j-\frac{1}{2}\right)\right)^i-\left(a_l+\epsilon_2\left(j-\frac{1}{2}\right)\right)^i\Bigg].
\end{align}
In particular, setting $n=3$ in \eqref{eq:Trphi} and comparing with \eqref{eq:energy} leads to the desired relation
\begin{equation}
 \text{Tr}\Phi_{\overrightarrow{\lambda}}^{3}=\sum_{l=1}^{N}a_l^{3}-\frac{3}{\epsilon_2}E_{\overrightarrow{\lambda}}^{(3),+}(g)+3\sum_{l=1}^{N}|\lambda_l|\Big[\frac{1+\epsilon_2^2}{\epsilon_2}(N+1-2l)-2a_l\Big].
\end{equation}
The last piece has to vanish, thus fixing the special value $a_l^*$
\begin{equation}
 a_l^*=\frac{1+\epsilon_2^2}{\epsilon_2}\frac{1}{2}(N+1-2l)=Q \rho_l,
\end{equation} 
where $\rho_l$ are the components\footnote[2]{In the basis $e_l,\;l=1\ldots,N$, vectors of $\mathfrak{h}^*=\mathbb{R}^{N-1}$ 
satisfying $\sum_{l=1}^{N}e_l=0$.} of the Weyl vector for $SU(N)$. 

Finally, the key relation connecting the operator $\text{Tr}\Phi^3$ and the energy of BO$_3$ integrable system is
\begin{equation}
\left.\text{Tr}\Phi_{\overrightarrow{\lambda}}^3\right|_{a_l^*}=\sum_{l=1}^{N}\left(a_l^*\right)^3-\frac{3}{\epsilon_2}E_{\overrightarrow{\lambda}}^{(3),+}(g).
\end{equation}


\bibliography{ref}
\bibliographystyle{JHEP-2}

  
\end{document}